\documentclass[useAMS,usenatbib]{mn2e}
\usepackage{amsmath,graphicx,txfonts}
\usepackage{deluxetable}
\usepackage{threeparttable}
\usepackage{natbib}
\usepackage{rotating}
\newcommand{\mr}{\mathrm}

\newcommand{\be}{\begin{equation}}
\newcommand{\ee}{\end{equation}}

\newcommand{\beq}{\begin{eqnarray}}
\newcommand{\eeq}{\end{eqnarray}}
\newcommand{\ensav}[1]{\left\langle #1 \right\rangle}

\def\aap{A\&A}
\def\apj{ApJ}
\def\apjl{ApJL}
\def\apjs{ApJS}
\def\mnras{MNRAS}
\def\aj{AJ}

\def\pasp{PASP}

\begin{document}

\title[HOD Modeling of Green Valley Galaxies]{Halo Occupation Distribution Modeling of Green Valley Galaxies} 
\author[E. Krause et al.]{Elisabeth Krause$^{1,2}$, Christopher M. Hirata$^1$, Christopher Martin$^3$, James D. Neill$^3$\& Ted K. Wyder$^3$\\
$^1$ California Institute of Technology, Department of Astronomy, MC 249-17, Pasadena, CA 91125, USA\\
$^2$ University of Pennsylvania, Department of Physics and Astronomy, Philadelphia, PA 19104, USA\\
$^3$ California Institute of Technology, Department of Astronomy, MC 290-17, Pasadena, CA 91125, USA}
\date{}
\pagerange{\pageref{firstpage}--\pageref{lastpage}} \pubyear{2012}
\maketitle
\label{firstpage}
\begin{abstract}
We present a clustering analysis of near ultraviolet (NUV) - optical color selected luminosity bin samples of green valley galaxies. These galaxy samples are constructed by matching the Sloan Digital Sky Survey Data Release 7 with the latest Galaxy Evolution Explorer source catalog which provides $NUV$ photometry. We present cross-correlation function measurements and determine the halo occupation distribution of green valley galaxies using a new multiple tracer analysis technique.

We extend the halo-occupation formalism, which describes the relation between galaxies and halo mass in terms of the probability $P(N, M_{\mr h})$ that a halo of given mass $M_{\mr h}$ contains $N$ galaxies, to model the cross-correlation function between a galaxy sample of interest and multiple tracer populations simultaneously. This method can be applied to commonly used luminosity threshold samples as well as to color and luminosity bin selected galaxy samples, and improves the accuracy of clustering analyses for sparse galaxy populations. 

We confirm the previously observed trend that red galaxies reside in more massive halos and are more likely to be satellite galaxies than average galaxies of similar luminosity. While the change in central galaxy host mass as a function of color is only weakly constrained, the satellite fraction and characteristic halo masses of green satellite galaxies are found to be intermediate between those of blue and red satellite galaxies.

\end{abstract}
\section{Introduction}
Most nearby galaxies fall into one of two well-known and well-characterized categories. They are either passively evolving elliptical galaxies with old stellar populations, red in color and typically living in high-density regions, or they are actively star-forming spiral galaxies with blue color. The latter often are field galaxies or reside in other low-density regions like cluster outskirts.

This blue/red galaxy color bimodality has been observed to be in place already around $z\sim 1$. The fraction of red galaxies increases with time \citep[e.g.,][]{F07} and therefore galaxies must transition from blue to red. Galaxies in this transitional stage characteristically show low levels of recent star formation. As ultraviolet emission is a sensitive tracer of recent star formation, these transition galaxies are easily identified in a $(NUV-r)$--$M_r$ color--magnitude diagram where they populate a ``green valley" between well-localized red and blue sequences \citep{W07}.

The relation between galaxy color and environment density also evolves with redshift, such that the fraction of red galaxies increases with time in dense environments but stays nearly constant for field galaxies \citep[e.g.,][and references therein]{C07}. This indicates the transition from blue to red galaxies may be driven by environmental processes, associated with the infall of a galaxy into a larger halo (``cluster"). Proposed mechanisms broadly fall into one of the following categories: galaxy--galaxy interactions, such as galaxy mergers, merger driven nuclear activity and high speed galaxy interactions, galaxy--intra cluster medium interactions (e.g., ram pressure stripping or thermal evaporation), and interactions between an infalling galaxy and the cluster potential (e.g., truncation through tidal forces). Observationally these are disentangled through their characteristic timescales, the dependence of their respective efficiencies on halo mass, and position within the cluster \citep[][]{T03,C06, ME07}; for example, galaxy mergers are expected to be one of the dominant processes in group-scale halos and in the outskirts of massive clusters. 

In the framework of $\Lambda$ cold dark matter (CDM) cosmology, the evolution and spatial distribution of dark matter halos is relatively well understood. A common technique for inferring the masses of halos hosting different galaxy populations is to measure the angular or spatial clustering of galaxies and relate it to the predicted clustering and abundance of dark matter halos. While the relation between galaxy and dark matter clustering on large scales can be approximately described by scale-independent biasing, the situation is more complicated -- and more informative about the physical processes at work -- on small scales: At the level of individual halos, so-called halo-occupation distribution (HOD) models \citep[e.g.,][]{BW02} describe the relation between galaxies and mass in terms of the probability that a halo of given mass contains $N$ galaxies of a given type. Then galaxy clustering, for example the two-point correlation function, is modeled as the sum of contributions from galaxy pairs residing in the same halo and from galaxy pairs living in different halos.

This method of interpreting galaxy correlation functions has been used extensively: For example, \citet[][see references therein for previous/high-z studies]{Zehavi10} analyze the completed (DR7) Sloan Digital Sky Survey (SDSS), and find, in agreement with previous results, that at the amplitude of the correlation function increases with luminosity, and that at fixed luminosity redder galaxies are more strongly clustered, due to redder galaxies being satellites in more massive (and thus more biased) halos.
Based on correlation function measurements over the redshift range $0.2<z<1.2$ from the Canada-France-Hawaii Telescope Legacy Survey, \citet{Coupon12} also find red central galaxies to reside in more massive halos than average central galaxies in the same luminosity sample.

The clustering of $(NUV-r)$ color selected galaxies from the \emph{Galaxy Evolution Explorer (GALEX)} survey has previously been studied by \citet{H07}, who measure the angular correlation function; \citet{H09} and \citet{L10} analyze spatial clustering as a function of star formation history and color respectively. These authors find the clustering of green galaxies to have intermediate strength compared to blue and red galaxies and to have a scale dependence closer to that of red galaxies. At small scales their analysis is strongly limited by statistics due to the small number density of green valley galaxies, limiting their ability to constrain the 1-halo term.

We extend the HOD formalism to simultaneously model the cross-correlation functions (CCF) of a sparse luminosity bin galaxy sample with multiple more abundant galaxy populations to study the environment of local green valley galaxies. We consider luminosity bin samples of green valley galaxies as the physical mechanisms populating the green valley, i.e., quenching star formation in blue galaxies or rejuvenating red galaxies, may depend on halo mass and thus vary with galaxy luminosity. Compared to an autocorrelation function based clustering analysis, measuring the CCF between (sparse) \emph{GALEX} selected galaxies and more abundant samples of SDSS galaxies reduces the shot noise contribution to our measurements, and also increases the effective volume probed beyond the combined \emph{GALEX}-SDSS footprint\footnote{We note that the increase in effective volume is limited to those regions of the SDSS footprint that are closer to the combined \emph{GALEX}-SDSS footprint than the largest scales probed by the CCF. Due to the patchy geometry of the \emph{GALEX} footprint, these regions cover nearly the entire SDSS footprint.}. Extending previous work on HOD models for CCFs \citep[e.g.,][]{K10} to simultaneously fit the clustering of the galaxy sample of interest with respect to multiple tracer populations is particularly helpful for analyzing the clustering of luminosity bin samples, which are harder to constrain than the more frequently used luminosity threshold samples.This allows us to put the separate piece of information found by \citet{H09} and \citet{L10} into a coherent analysis including HOD modeling, and improve the statistics due to the larger survey area included in the newest data release. 

Throughout this analysis we assume a flat $\Lambda$CDM cosmology with $\Omega_{\mr m}  =0.3$ and $\sigma_8  = 0.8$. Unless specified otherwise, all distances are coming and quoted in $\mr{Mpc}/h$, and all absolute magnitude are given in $h=1$ units.
\section{Data}
\subsection{SDSS}
The Sloan Digital Sky Survey \citep{SDSS} mapped most of the high-latitude sky in the northern Galactic cap using a dedicated wide-field 2.5 m telescope at Apache Point Observatory \citep{Gunn06} with the SDSS camera \citep{Gunn98}. The raw imaging data were processed by a series of pipelines performing photometric calibration \citep{Hogg01, Ivezic04, Tucker06}, photometric reduction \citep{Lupton01}, and astrometric calibraton \citep{Pier03}.
Data release 7 \citep[DR7][]{SDSS7} of the  
spectroscopic sample provides (u'g'r'i'z')-photometry \citep{Fukugita96, Smith02}
and spectra for nearly $900000$ galaxies with $m_{\mr r} <17.77$ over $8000$ square degrees. These galaxies were selected from the photometric survey for spectroscopic follow-up using specific algorithms for the main galaxy sample \citep{Strauss02} and luminous red galaxies \citep{Eisenstein01}. The main spectroscopic galaxy sample is nearly complete to $r<17.77$ and has a median redshift of $z\sim 0.1$. Based on these observations, the NYU Value Added Galaxy Catalog \citep[VAGC,][]{NYUVA} contains galaxy samples which have been constructed for large-scale structure studies: all magnitudes are re-calibrated \citep{Padmanabhan08} and K-corrected \citep{kcorrect}, and the radial selection function and angular completeness are carefully determined from the data. We restrict this sample to $m_{\mr r} <17.6$ to ensure uniform completeness of faint galaxies across the survey area.

Due to fiber placement in the SDSS spectrograph \citep{tiling}, galaxies closer than $55''$ cannot be observed on the same spectroscopic plate, and hence no redshifts have been measured for about $7\%$ of all targeted galaxies . The lack of observed close galaxy pairs affects the measured correlation functions on  small scales. While it is possible to correct for fiber collisions down to $0.01\, \mr{Mpc}/h$ \citep{L06}, the number density of green valley galaxies is too small to obtain correlation function measurements at such small separations and we simply assign galaxies with missing spectra the redshift of its nearest neighbor. This method has been shown to work well for projected correlation functions above the scale corresponding to $55''$ \citep{Z05}. For the most distant galaxies in our sample the fiber collision scale is $0.07$ comoving $\mr{Mpc}/h$ and we measure correlation functions only on perpendicular scales $r_{\mr p} \ge 0.1 \, \mr{Mpc}/h$.

Spectral line measurements and mass estimates for these galaxies are taken from the MPA-JHU catalog.\footnote{http://www.strw.leidenuniv.nl/~jarle/SDSS/} We use the former to classify the $(NUV-r)$ selected transitional galaxies with emission line diagrams and to compare $(NUV-r)$ color selection with spectroscopic separation of active and quenched galaxies based on $D_{\mr n}4000$ (Fig.~\ref{fig:D4000}). Note that these quantities are estimated from a fiber size of $3''$, and due to low redshift of our galaxy sample these measurements may not be representative of the luminosity averaged properties of a galaxy but rather be dominated by central (bulge dominated) regions.

\subsection{GALEX}
$NUV$ photometry for this project is taken from the GALEX Medium Imaging Survey (MIS) Source
Catalog (GMSC, Seibert et al. in prep.) derived from the GALEX GR6 data release, which
provides unique measurements of point and extended sources
up to 1 arcminute diameter in the GALEX bands (Seibert at al., in prep.). The $NUV$ source catalog covers 4827 square degree at $\lambda_{\mr{eff}} = 2316\AA$ with a resolution of $5.3''$ and reaching a depth $\approx$ 23 mag.

GALEX  has a circular field of view of $1.2^\circ$ which is sampled at $1.5''$. Each field targets a pre-defined position on the sky, resulting in a hexagonal tiling of the survey. These angular selection parameters are contained in exposure time,
coverage and flag maps in $\mathtt{HEALpix}$ \citep{HEALpix} format accompanying the GMSC, which we use to define the combined footprint and select our galaxy sample as detailed in section \ref{sec:overlap}.
\subsection{SDSS-MIS Cross-Match}
\label{sec:overlap}
\begin{figure}
\includegraphics[width = 0.25\textwidth, angle = 90, trim  = 0mm 0mm 0mm 0mm, clip = true]{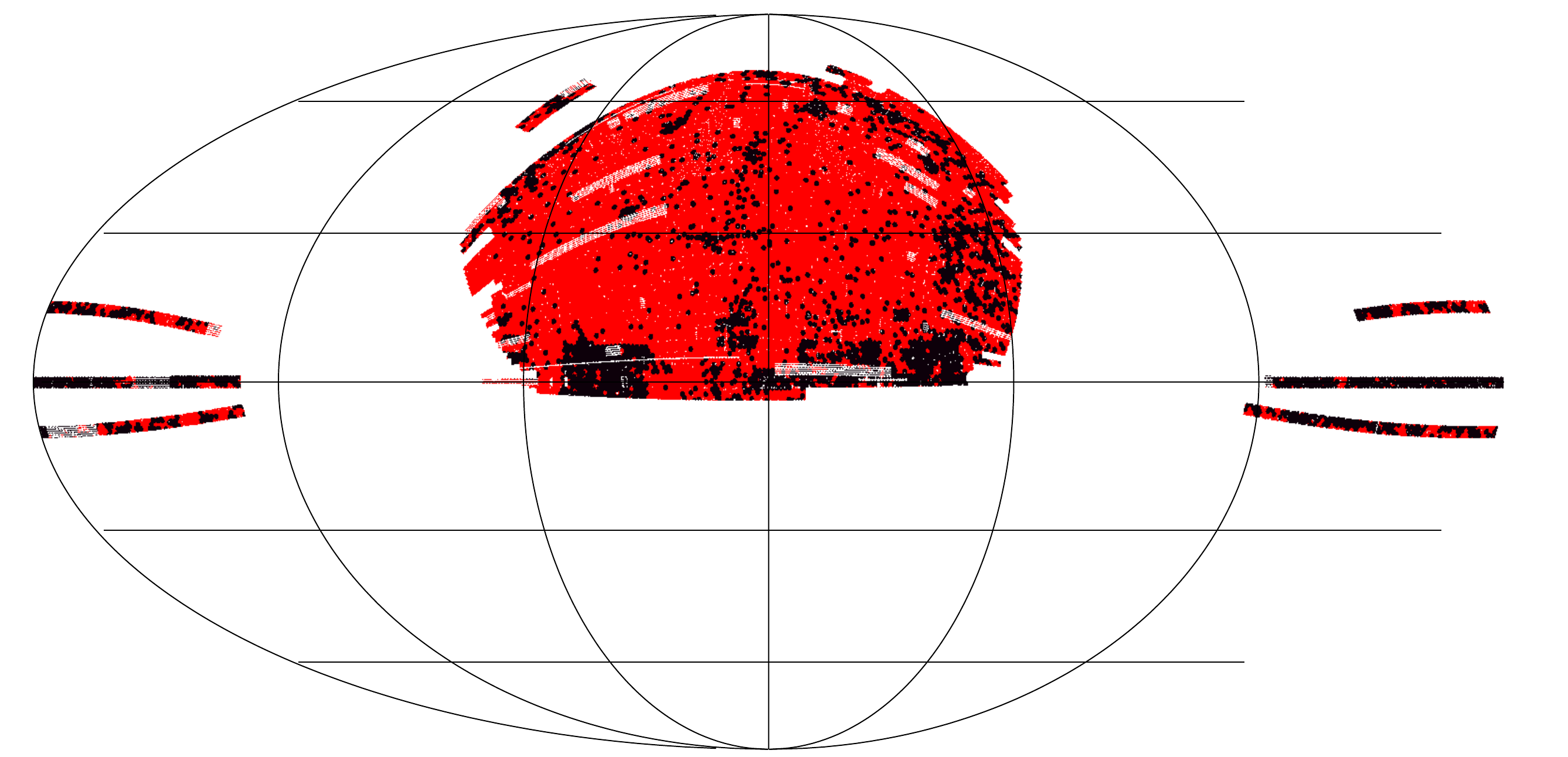}
\caption[Combined survey footprint]{Combined SDSS + GALEX MIS footprint. The area covered by the VAGC at an angular completeness $f_{\mr{comp}} > 0.7$ is shown in red, the final overlap area of 1945 square degrees between VAGC and MIS, as detailed in section \ref{sec:overlap}, is shown in black.}
\label{fig:area}
\end{figure}

In order to match the VAGC with $NUV$ detections, we first construct the combined footprint of these two surveys. This is done by converting the VAGC angular selection function, which is given in terms of $\mathtt{Mangle}$ polygons \citep{Mangle1}, into the pixelized $\mathtt{HEALpix}$ format \citep{Mangle2}. Then we multiply the angular selection functions of the VAGC and MIS in each pixel (at resolution $N_{\mr{side}}=2048$) and restrict the overlap region to pixels where the angular completeness fraction of both surveys is larger than $0.7$. This results in a combined survey with an effective area of 2708 square degrees. Furthermore, we require tiles to have $NUV$ exposure times $t>1000$ s, which reduces the combined effective area to 1945 square degrees. This final overlap region is shown in black in Fig.~\ref{fig:area}.

\begin{figure}
\includegraphics[width = 0.5\textwidth]{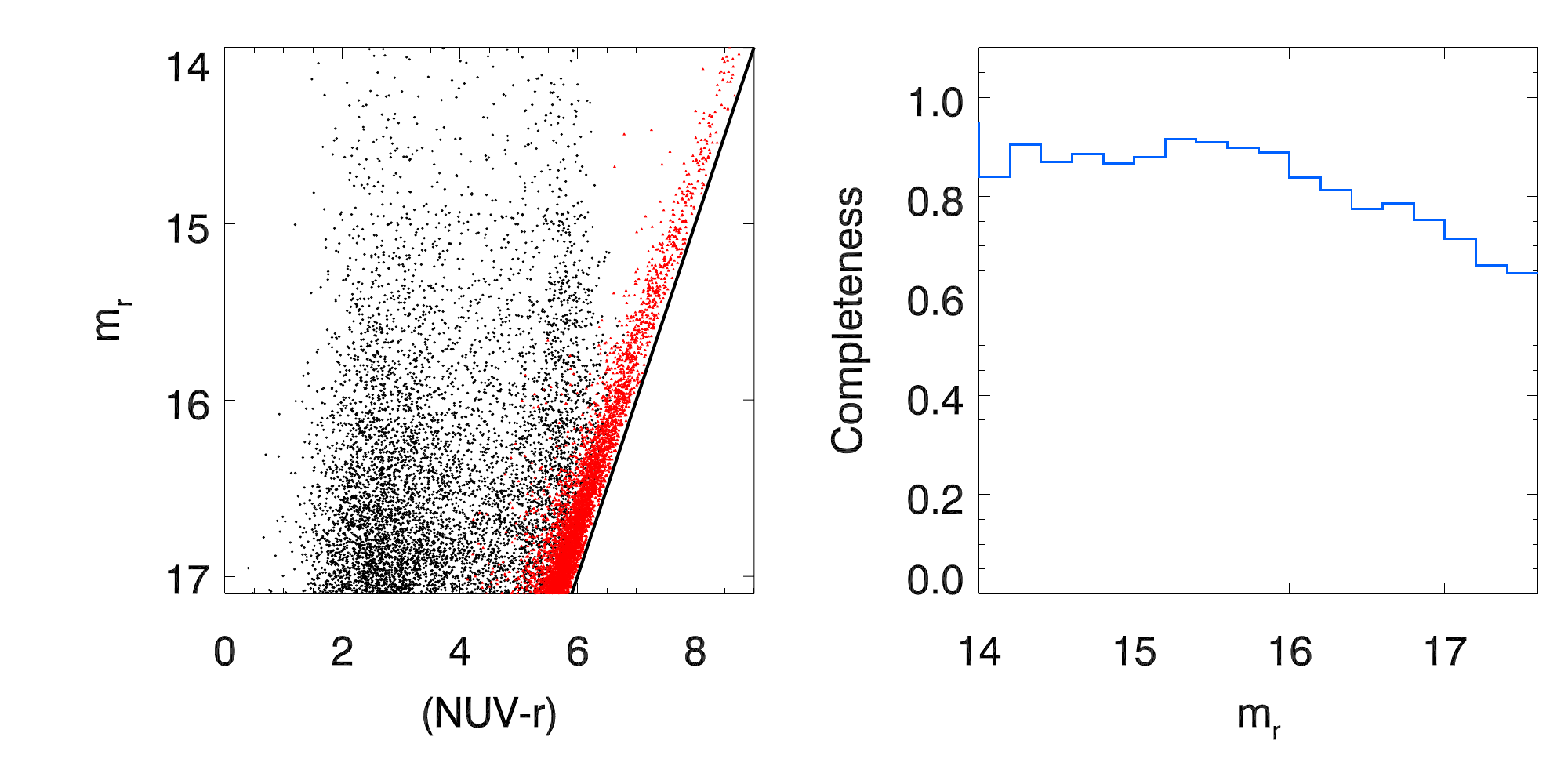}
\caption[Completeness of the cross-match sample]{Completeness of the cross-match sample.\emph{Left:} Apparent magnitude--$(NUV-r)$ color diagram. Black dots show a random subset of VAGC galaxies with $NUV$ cross-match. Red dots indicate VAGC galaxies without $NUV$ detections, which have been placed at the detection limit $NUV= 23$ and corrected for position dependent galactic extinction.\newline
\emph{Right:} Completeness of the $NUV$ cross-match as a function of apparent magnitude.}
\label{fig:complete}
\end{figure}

We cross-match all galaxies in the VAGC within this overlap area with $NUV$ detections using a search radius of $4''$. In order to construct a complete statistical sample, we then restrict the cross-match with various cuts summarized in table~\ref{tab:limits}. Due to deblending and centering issues for nearby or very bright objects, the $NUV$ and $r$ band photometry pipelines may report positions for these objects that are farther separated than the matching radius, leading to spurious non-detections. Furthermore, the astrometric and photometric precision of the GALEX detections declines toward the edges of each tile, and near light echos and other imaging artifacts and we exclude this regions as detailed in table~\ref{tab:limits}. The color--apparent magnitude distribution and completeness of the final cross-match sample is shown in Fig.~\ref{fig:complete}. For apparently bright galaxies ($m_{\mr r} \lesssim 16$) the blue sequence (around $(NUV-r) \approx2-3$) and the red sequence (around $(NUV-r) \approx5-6$) are clearly visible. No galaxies are found with $(NUV-r)\gtrsim6.5$ though these should well be within the GALEX detection limit (indicated by the inclined line) at these magnitudes if they existed. For these bright galaxies far from the $NUV$ detection limit the cross-match completeness is around $90\%$, it decreases for fainter objects as the $NUV$ detection limit moves into the color-magnitude space occupied by red galaxies. In order to retain a nearly complete sample of green valley galaxies we cut the cross-match sample at $m_{\rm r} < 17.1$. The resulting cross-match catalog has a completeness of $76\%$, i.e. $76\%$ of galaxies in the VAGC catalog, that meet the magnitude and redshift criteria described above are at a position with GALEX coverage as detailed in table~\ref{tab:limits}, have a $m_{NUV}<23.0$ GALEX detection.

Finally, we use $\mathtt{kcorrect v4.2}$ \citep{kNUV} to calculate absolute $\mr{NUV}_{0.1}$ magnitudes of the cross-match galaxies k-corrected to $z=0.1$. As the redshift evolution in the $NUV$ is not very well constrained, we do not attempt to apply evolution corrections to the $NUV$ nor optical magnitudes. Similarly, we do not attempt to correct the $(NUV-r)$ colors for intrinsic extinction. To isolate transitional galaxies and avoid identifying dusty (edge on) spiral galaxies as green valley objects, we only consider objects with $r$-band isophotal axis ratio $b/a >0.5$.

\begin{table}
\caption{Cross-match sample definition$^a$}
\begin{threeparttable}
\begin{tabular}{l r}
\hline
Parameter & Limits\\
\hline
r-band magnitude& $14.1 < r< 17.1$\\
redshift & $0.02< z <0.2$\\
\emph{GALEX} field radius & $fov_-radius <0^\circ.55$\\
\emph{GALEX} exposure time& $t>1000$ s\\
$NUV$ flag & $nuv_-artifact \leq 1$\\
$NUV$ magnitude & $16.0<\mr{NUV}<23.0$\\
SDSS/ $NUV$ angular completeness & $f_{\mr{comp}} > 0.7$\\
\hline
\end{tabular}
\begin{tablenotes}
\item $^a$The parent catalog is the NYU VAGC $\mathtt{dr72bright}$.
\end{tablenotes}
\end{threeparttable}
\label{tab:limits}
\end{table}%

\section{Sample Definition}
\begin{figure}
\includegraphics[width = 0.5\textwidth]{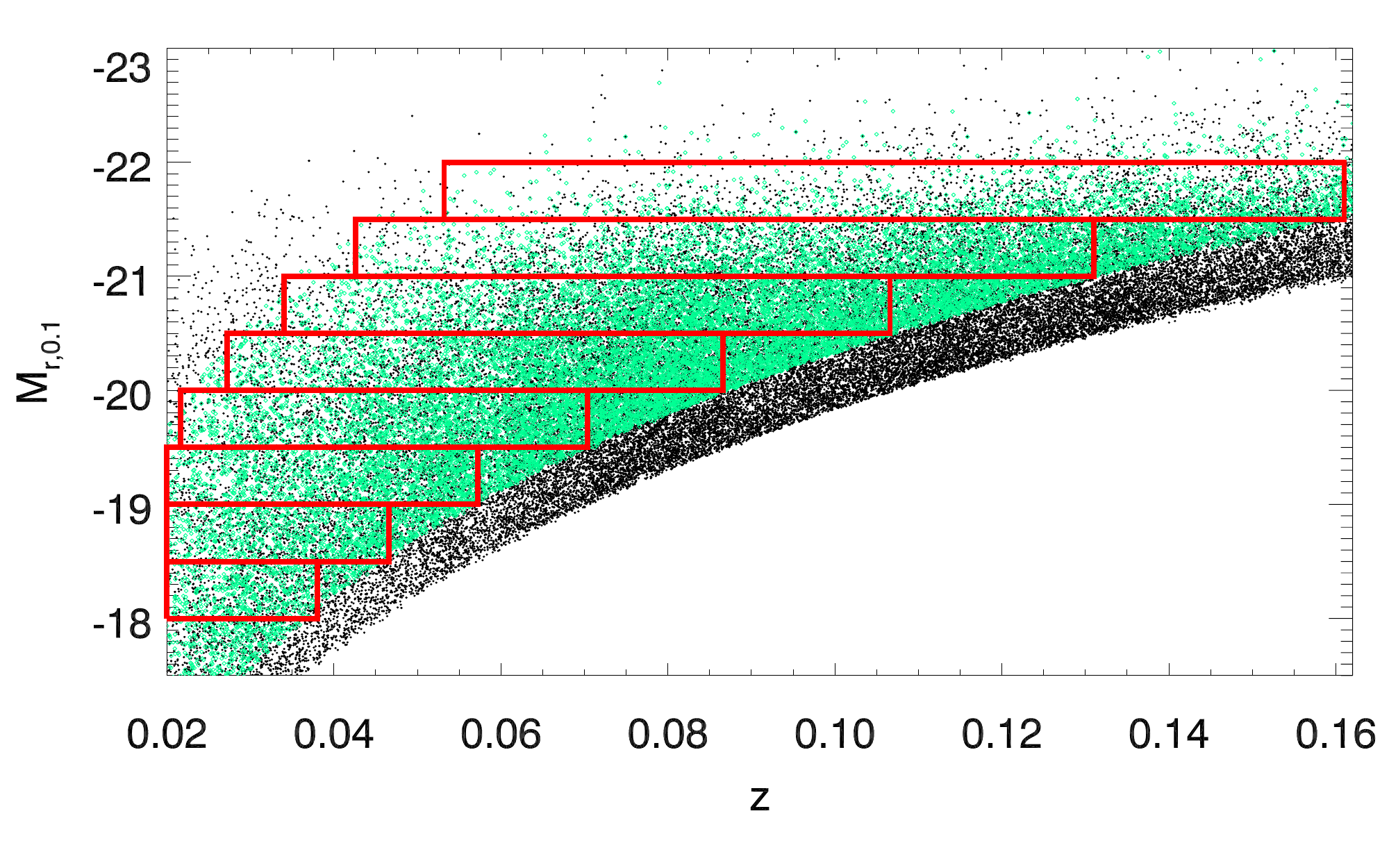}
\caption[Definition of volume-limited galaxy samples]{Volume-limited color selected galaxy samples: Black dots show a random subsample of VAGC galaxies with $m_{\mr r} < 17.6$, subsampled by a factor 10. Green symbols indicate green valley galaxies identified based on their $(NUV-r)$ color, which are restricted to $14.1 < m_{\mr r} < 17.1$ to ensure (near) completeness of the cross-matched sample. Red boxes indicate the location of volume-limited color selected galaxy samples in luminosity redshift space.}
\label{fig:zMr}
\end{figure}
\begin{figure}
\includegraphics[width = 0.5\textwidth]{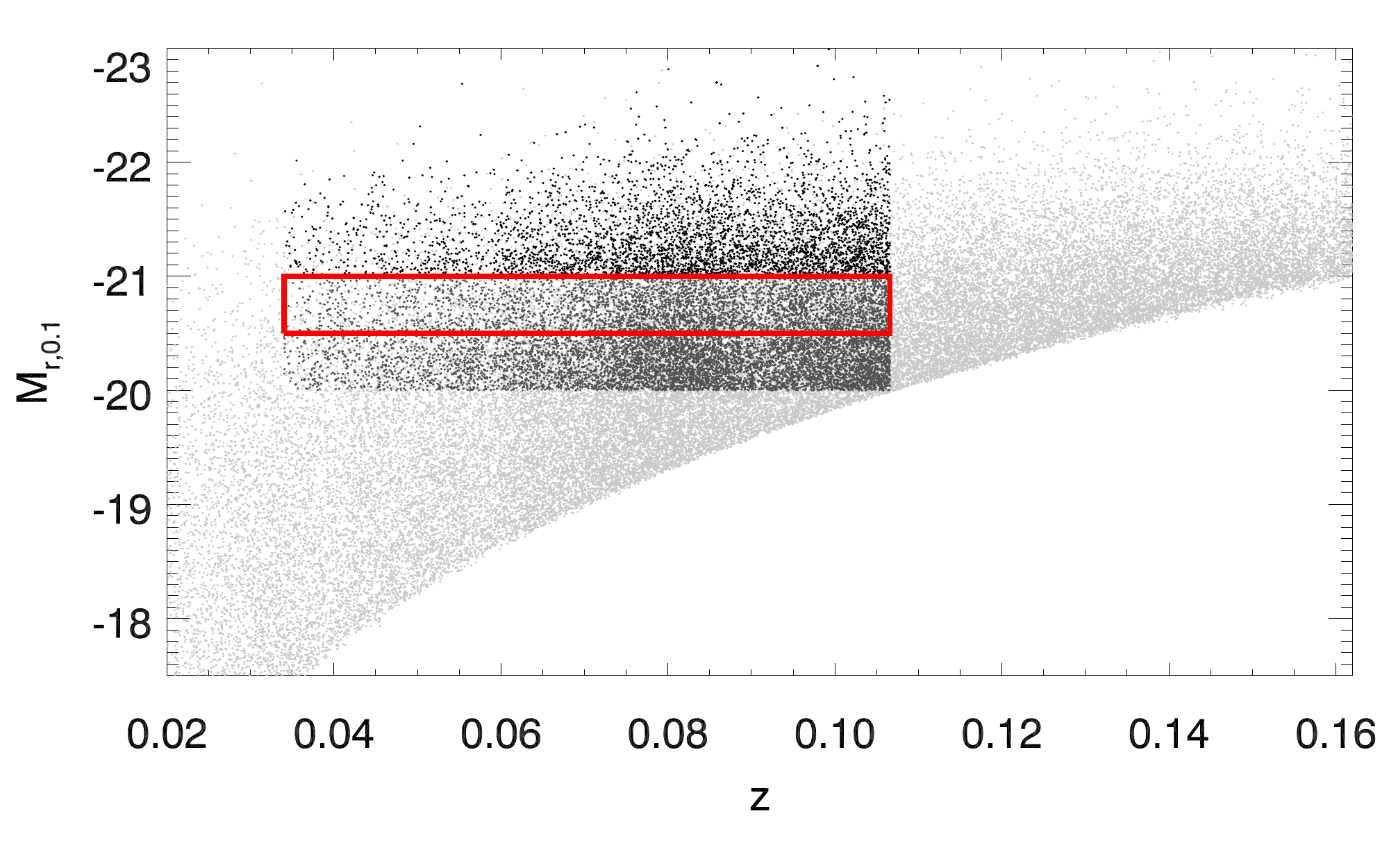}
\caption[Definition of volume-limited galaxy samples]{Definition of volume-limited SDSS galaxy reference samples: Light-gray dots show a random subsample of VAGC galaxies with $m_{\mr r} < 17.6$, subsampled by a factor 10. The red box indicates the location of $[-20.5,-21]$ magnitude range volume-limited color selected galaxy samples. The dark gray points show the extend of the volume limited ``faint'' SDSS galaxy reference sample associated with this color selected galaxy sample, the black dots illustrate the associated ``bright'' luminosity threshold reference sample. The definitions are analogous for other magnitude ranges, hence we show only one example to improve clarity of the plot.}
\label{fig:zMr_fb}
\end{figure}

In order to work with well-defined galaxy populations, we construct a number of volume-limited samples. As the properties of green valley galaxies may vary with luminosity, we define samples of width 0.5 in absolute magnitude, and find the redshift range over which all galaxies in this sample have apparent magnitudes $14.1 < m_{\mr r} < 17.1$ (the magnitude range of the cross-matched catalog), c.f. Fig.~\ref{fig:zMr}. The VAGC has less stringent apparent magnitude requirements ($10 < m_{\mr r} < 17.6$), and we define two samples of SDSS galaxies occupying the same volume as each luminosity bin sample of $NUV$ detected objects, which are used for the cross-correlation analysis. These samples are described in detail in table~\ref{tab:sample1}. Specifically, for the luminosity bin $[M_{\mr{r,min}}, M_{\mr{r,max}}]$ we define the ``bright'' sample of SDSS galaxies to contain all galaxies in the same redshift range brighter than $M_{\mr{r,max}}$, and the ``faint'' sample to consist of the volume-limited sample $[M_{\mr{r,min}}+0.5, M_{\mr{r,max}}]$. The definition of these samples luminosity redshift space is illustrated in Fig.~\ref{fig:zMr_fb}. We refer to the union of these two samples, which is a luminosity threshold sample with threshold $M_{\mr{r,min}}+0.5$, as the SDSS ``all'' sample. 

\begin{deluxetable}{cr rrr||r r}
\tablecolumns{7}
\tablewidth{0pt}
\tablecaption{Volume-limited galaxy samples}
\tablehead{
\multicolumn{4}{c}{Green Valley sample}&\colhead{}&\multicolumn{2}{c}{SDSS samples}\\
\cline{1-4} \cline{6-7}\\
\colhead{$M_{\mr r}$} &\colhead{$\ensav{z}$}&\colhead{$N_{\mr G}$}&\colhead{$\bar{n}_{\mr G}$}&\colhead{}&\colhead{$N_{\mr f}$} & \colhead{$N_{\mr b}$}}
\startdata
$[-18,-18.5]$ & 0.031& 285&1.09 
&&15714& 22177\\
$[-18.5,-19]$ & 0.036& 595& 1.19 
& &24725     &  28488\\
$[-19,-19.5]$ & 0.044& 869& 0.92
& &38537    &   33041\\
$[-19.5,-20]$ & 0.055& 1191& 0.67
& &62193  & 37310 \\
$[-20,-20.5]$ & 0.068&1746&  0.54
& &95204 & 36561\\
$[-20.5,-21]$ & 0.083&2028& 0.35
& &109490 &  23586\\
$[-21,-21.5]$ & 0.102& 1383&  0.13 
& &112647 &12073 \\
$[-21.5,-22]$ &0.128 &775&  0.04
& &87676 & 4458 \\
\enddata
\tablecomments{The first two columns give the magnitude range $[M_{\mr{r,min}}, M_{\mr{r,max}}]$ and mean redshift of the green valley galaxy samples illustrated in Fig.~\ref{fig:zMr}. $N_\mr{G}$ is the number of green valley galaxies in this sample, and $\bar{n}_{\mr G}$ their mean comoving density per $10^{-3}\,(\mr{Mpc}/h)^3$. $N_{\mr f}$  and $N_{\mr b}$ are the number of SDSS galaxies in the faint and bright sample in the same volume; the bright sample consists of galaxies in the same volume that are brighter than $M_{\mr{r,max}}$, and the faint sample contains galaxies in the magnitude range $[M_{\mr{r,min}}+0.5, M_{\mr{r,max}}]$.}
\label{tab:sample1}
\end{deluxetable}%

\subsection{Finding the Green Valley}
\label{sec:GV}
We define the location of the green valley in $(NUV-r)$ color--magnitude space by fitting blue and red sequences to the color distribution of each volume-limited sample. We include galaxies without $NUV$ detections, which otherwise meet all cross-match criteria and are optically red ($(g-r)> 0.8$), by placing them at the $NUV$ detection threshold, correcting for position dependent galactic extinction and assigning the mean k-correction of cross-matched galaxies which are within $\Delta(NUV-r)=\pm0.1$ mag, $\Delta M_r= \pm 0.1$ mag, and $\Delta z = \pm 0.02$ of the unmatched galaxy. We the find the center and scatter of the color sequences by fitting each sequence with a Gaussian. Initially, we cut the distribution at $(NUV-r)$ = 4.2 and fit a Gaussian to each side. We then iteratively adjust the fitting range to include the galaxies within $1\sigma$ of the peak location on the ridge toward the Green valley. The best-fit parameters for each sample are shown in Fig.~\ref{fig:CM} along with fits to the blue and red sequence obtained by \citet{W07}, which are based on a different fitting scheme and one continuous galaxy sample weighted by the $v_{\mr{max}}$ method instead of using disjunct volume-limited samples. As we include $NUV$ non-detections, which are unaccounted for by \citet{W07}, our red sequence is slightly redder for faint galaxies, but otherwise these results agree very well.

\begin{figure*}
\includegraphics[width = 0.8\textwidth]{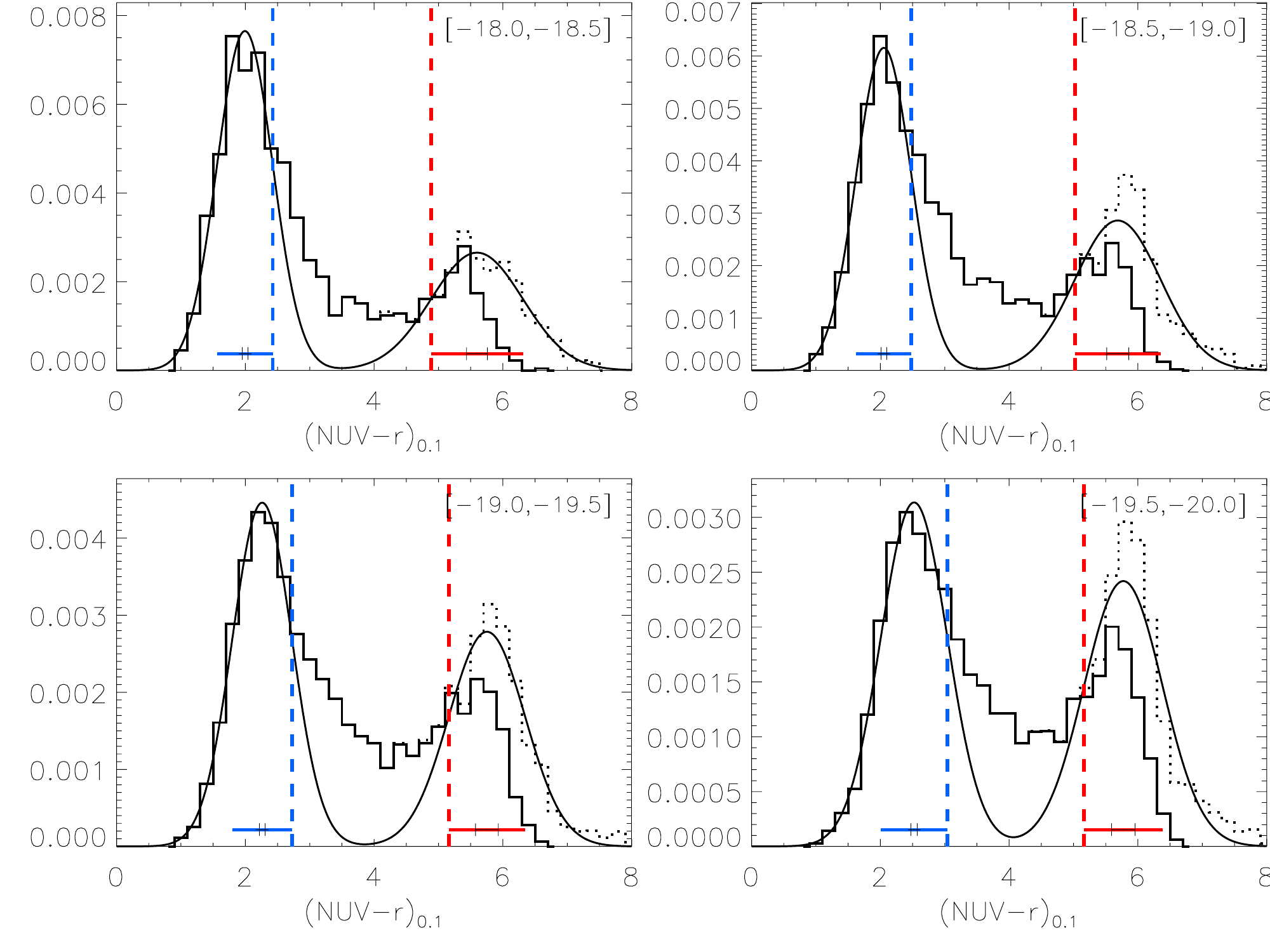}
\includegraphics[width = 0.8\textwidth]{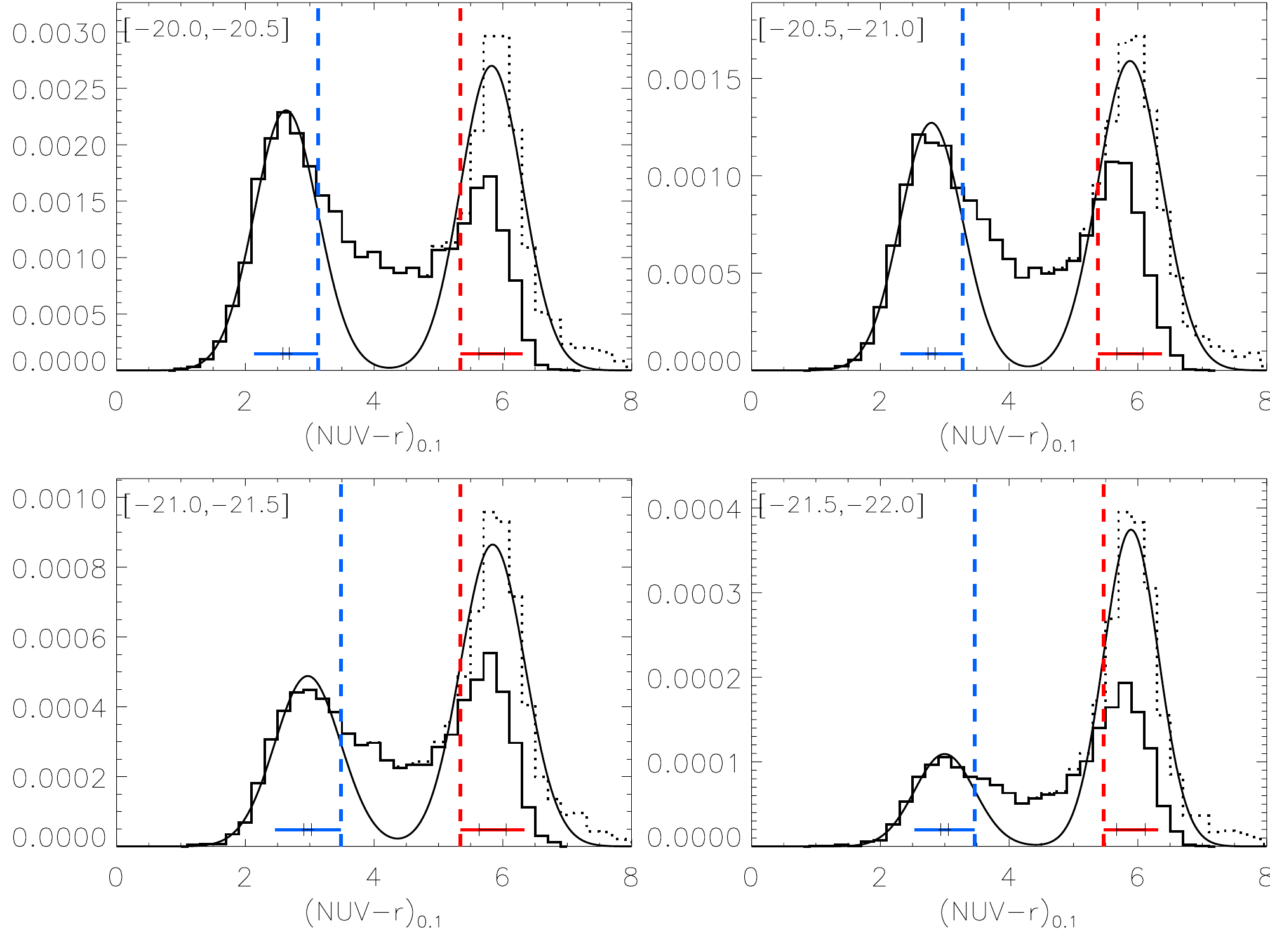}
\caption[Galaxy density as a function of $(NUV-r)$ color with fits to blue and red sequence]{Comoving density of the volume-limited galaxy samples as a function of $(NUV-r)$ color. Solid histograms show all $NUV$ detected galaxies. The dotted histograms include $NUV$ non-detections, which otherwise meet all cross-match criteria and are optically red ($(g-r)> 0.8$), placed at the $NUV$ detection threshold, corrected for position dependent galactic extinction and assigned the mean k-correction of cross matched galaxies which are within $\Delta(NUV-r)=\pm0.1$ mag, $\Delta M_r= \pm 0.1$ mag, and $\Delta z = \pm 0.02$ of the unmatched galaxy. The solid line shows the double Gaussian fit to the blue side of the blue sequence and the red side of the red sequence, as described in \ref{sec:GV}, and the vertical blue and red lines show the $1\sigma$ ridge of the color sequences derived from these fits. The colored error bars also indicate the $1\sigma$ scatter of the color sequences centered on their respective peak. The black error bars illustrate the mean photometric uncertainty in the $(NUV-r)$ color of blue/red galaxies.}
\label{fig:NUVr}
\end{figure*}

\begin{figure}
\includegraphics[width = 0.5\textwidth]{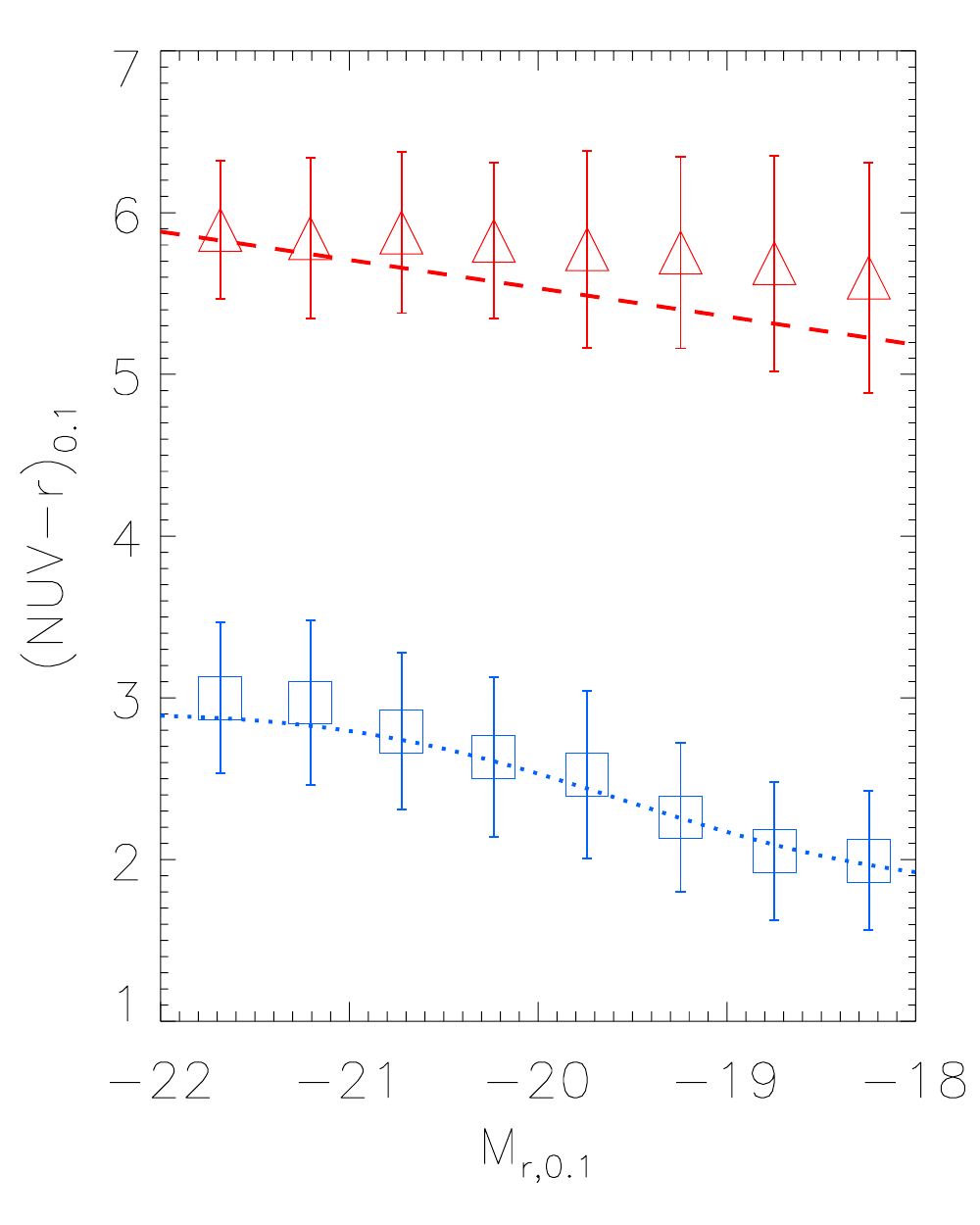}
\caption[Location and scatter of the blue and red sequence]{Defining the green valley: Symbols and error bars show the location and scatter of the blue and red sequence from the fits in Fig.~\ref{fig:NUVr}. Lines show the best-fit sequences from \citet{W07} transformed to our magnitude units.}
\label{fig:CM}
\end{figure}

The  black error bars in Fig.~\ref{fig:NUVr} illustrate the mean photometric uncertainty in the $(NUV-r)$ color of blue/red galaxies, suggesting that asymmetric scatter into the green valley due to photometric uncertainties is small compared to the intrinsic scatter of the red sequence. 

\subsection{Sample Properties}
\begin{figure}
\includegraphics[width = 0.5\textwidth]{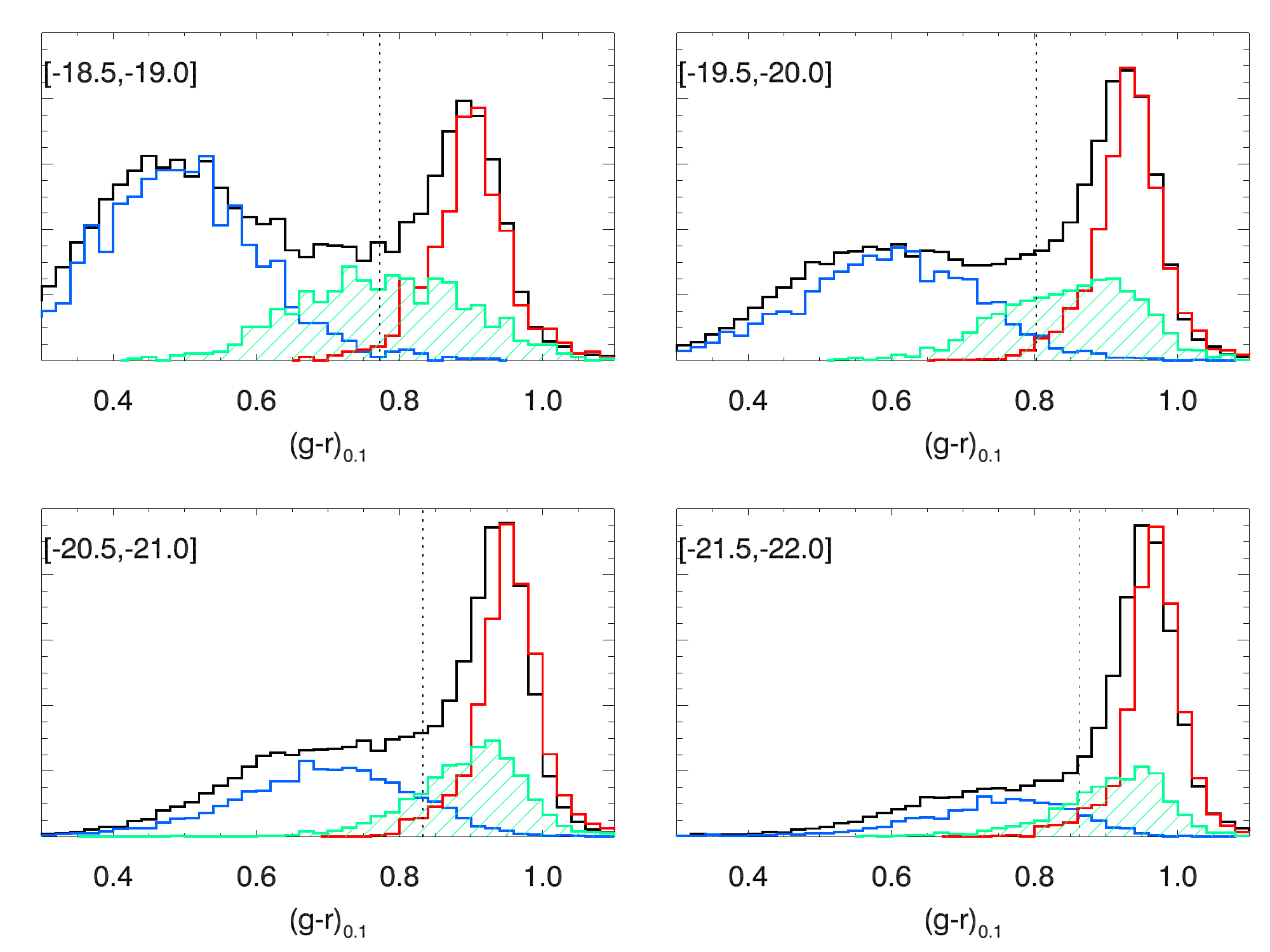}
\caption[Distribution of $(NUV-r)$ selected blue/green/red galaxies in $(g-r)$ space]{Colored histograms show the distribution of $(NUV-r)$ selected blue/green/red galaxies in $(g-r)$ space. The black histogram shows the distribution of all SDSS galaxies in the volume-limited sample, but not restricted to the combined footprint. The vertical line shows the color cut separating blue and red galaxies from \citet{Zehavi10}.}
\label{fig:nuvr_gr}
\end{figure}
\begin{figure}
\includegraphics[width = 0.5\textwidth]{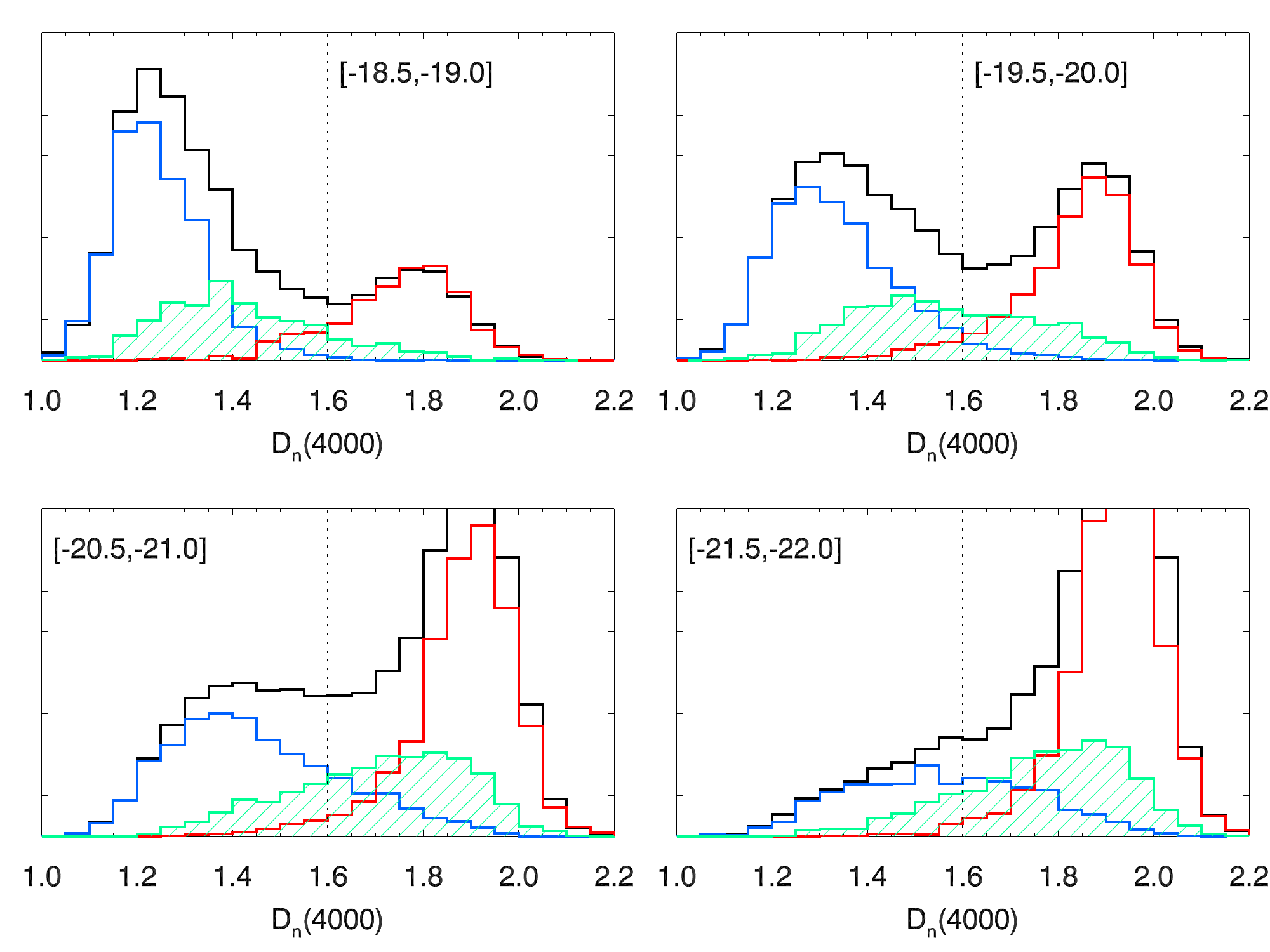}
\caption[Distribution of $(NUV-r)$ selected blue/green/red galaxies as a function of $\mr{D_n}4000$]{Same as Fig.~\ref{fig:nuvr_gr} but for $\mr{D_n}4000$. The vertical line shows the separation between quenched ($\mr{D_n}4000> 1.6$ and star forming galaxies used in \citet{Tinker11}.}
\label{fig:D4000}
\end{figure}
In order to facilitate the comparison with other studies of transitional galaxies based on optical criteria, we characterize the $(NUV-r)$ selected galaxies in other parameter spaces. 

Figure \ref{fig:nuvr_gr} and Fig.~\ref{fig:D4000} show the distribution of $(NUV-r)$ selected galaxies in $(g-r)$ color space and as a function of the Balmer break index $\mr{D_n}4000$. Here the red sample again includes NUV non-detections as described in section \ref{sec:GV}. The vertical lines indicate the transition between blue/red and star forming/quenched galaxies based on $(g-r)$ and $\mr{D_n}4000$ respectively. Most faint $(NUV-r)$ selected green valley galaxies are optically blue and and would be classified as star forming by both of these criteria. On the other end, a large fraction of luminous, $(NUV-r)$ selected transitional galaxies would be classified as red/quenched by both of these criteria. %

\begin{figure}
\includegraphics[width = 0.5\textwidth]{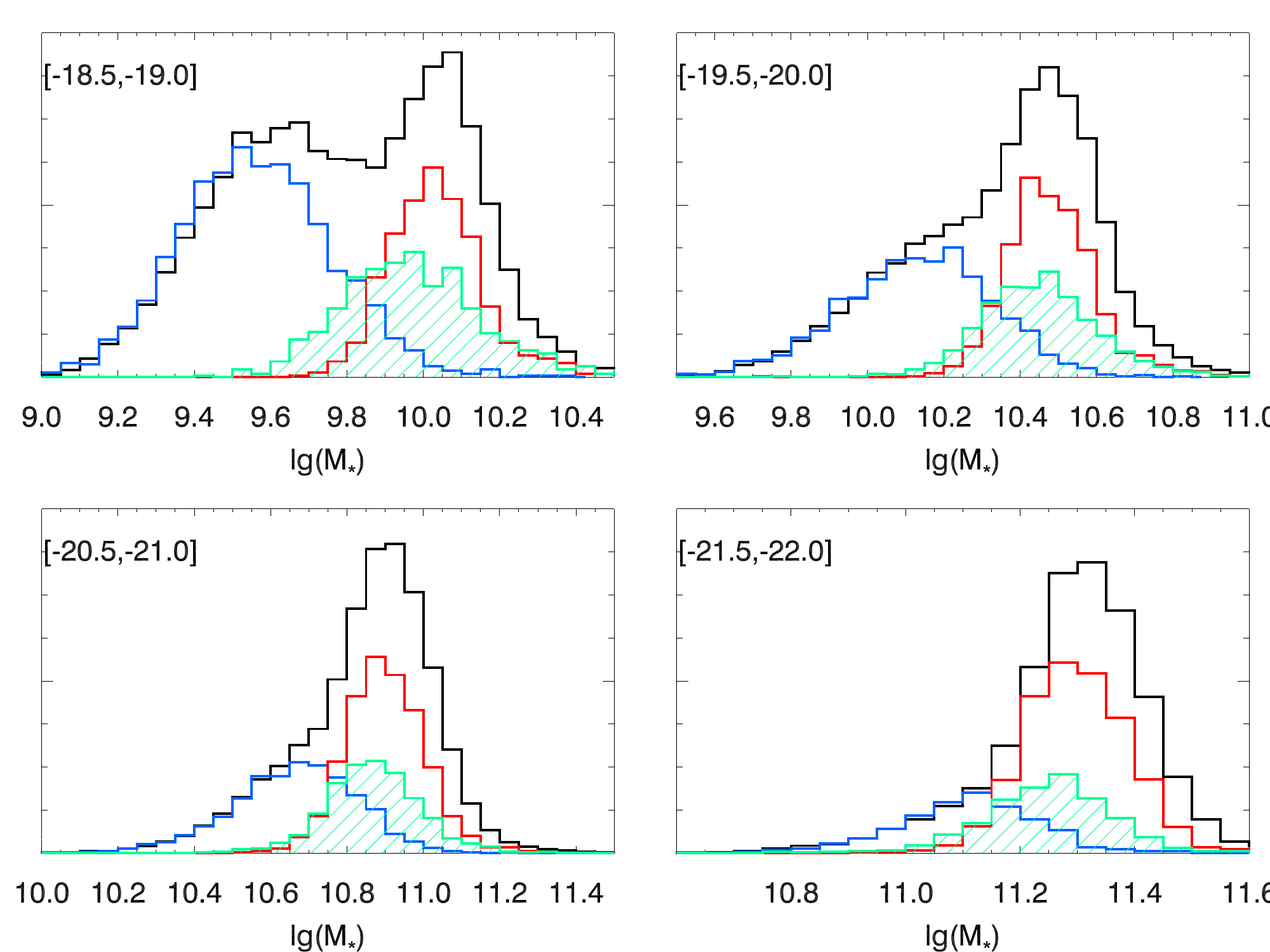}
\caption[Distribution of $(NUV-r)$ selected blue/green/red galaxies as a function of stellar mass]{Same as Fig.~\ref{fig:nuvr_gr} but for stellar mass.}
\label{fig:NUV_mass1}
\end{figure}

Furthermore,  Fig.~\ref{fig:NUV_mass1} shows the distribution of stellar masses as a function of $(NUV-r)$ color. The stellar masses are taken from the MPA-JHU catalog and are based on \citet{Kauffmann03}. At fixed luminosity, green valley galaxies and red sequence galaxies have similar stellar masses. 

\begin{figure}
\includegraphics[width = 0.5\textwidth]{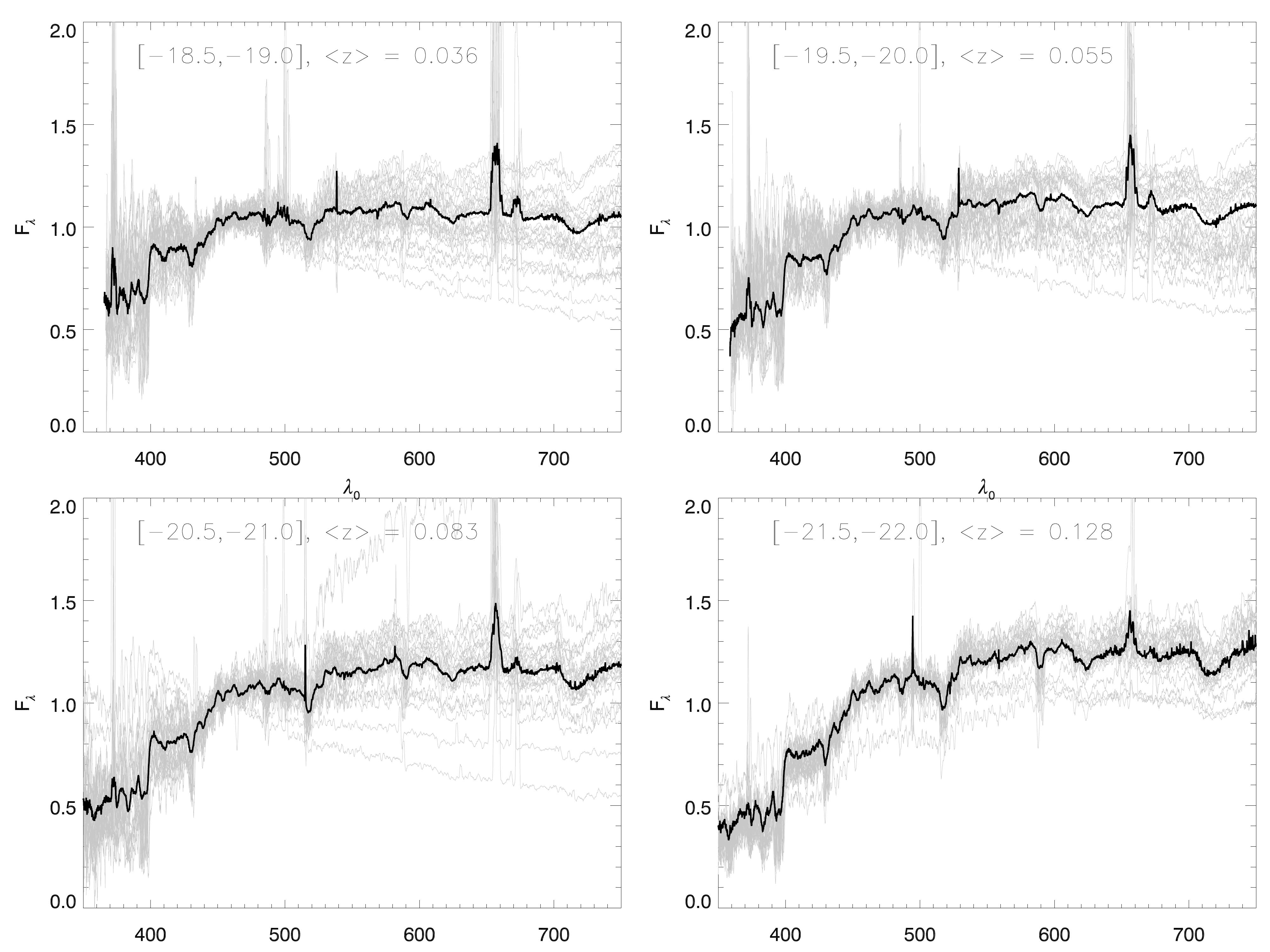}
\caption{Stacked spectra of (NUV-r) selected green valley galaxies for different luminosity bins as a function of restframe wavelength. The thin gray lines show 25 randomly chosen individual spectra, boxcar smoothed over 10 pixel to enhance readability.}
\label{fig:spectra}
\end{figure}

We illustrate the distribution of green valley galaxy spectra for different luminosity bins in FIg.~\ref{fig:spectra}. The thick line shows the mean spectrum obtained from stacking all green valley galaxies (with $r$-band isophotal axis ratio larger than 0.5) within $\Delta z=0.02$ of the mean redshift of each luminosity bin. The individual spectra are normalized to the median flux in the 410-500 nm range, giving each galaxy equal weight. The thin gray lines show smoothed individual spectra of 25 galaxies randomly chosen from those used in the stacking process. While we use the spectra mask to exclude pixels flagged by the SDSS spectra reduction pipeline, these spectra contain residual atmospheric [OI] and OH. Note that the fiber diameter of 3'' roughly corresponds 1.5 kpc/$h$ and $z=0.036$, to 3 kpc/$h$ at $z = 0.083$, and to 4.8 kpc/$h$ at $z =0.13$. The stacked spectra show that, on average, green valley galaxies have red bulges and some amount of AGN activity. All spectra show $\mr H_\alpha$, or a combination of $\mr H_\alpha$ and $[\mr{NII}]$, emission, which we classify further using emission line diagnostics in Tab.~\ref{tab:BPT}. For green valley galaxies with emission line measurements with $S/N > 3$ the AGN fraction is substantial, especially among the more luminous ones. Note that we use emission lines from the MPA-JHU with rescaled flux errors. However, in particular for the less luminous samples at lower redshifts there is considerable spread among objects, limiting the informative value of the stacked spectra.
\begin{table}
\caption{Classification of Green Valley Galaxies}
\begin{threeparttable}
\begin{tabular*}{0.45\textwidth}{cc ccccc}
\hline
$M_\mr{r}$ &$f^b_{\mr{H}_\alpha}$& $f^c_{\mr{SF}}$ & $f^c_{\mr{comp}}$ & $f^c_{\mr{AGN}}$&&$f^d_{\mr{low\,S/N}}$\\
\hline
$[-18,-18.5]$  &0.94& 0.56 &0.04  &0.04 &&0.34\\
$[-18.5,-19]$  &0.92&  0.47&0.14  & 0.06 &&0.34\\
$[-19,-19.5]$  &0.92&  0.31& 0.22 & 0.09&&0.38 \\
$[-19.5,-20]$  &0.90&  0.28&0.25  & 0.10 &&0.38\\
$[-20,-20.5]$  &0.84&  0.16&0.20  & 0.11&&0.55\\
$[-20.5,-21]$  &0.75&  0.12& 0.16 & 0.10&&0.66\\
$[-21,-21.5]$  &0.69&  0.09& 0.11 &  0.09&&0.72\\
$[-21.5,-22]$  &0.60&  0.06& 0.09 & 0.08&&0.81\\
\hline
\end{tabular*}
\begin{tablenotes}
\item $^b$ Fraction of green valley galaxies with $\mr{H}_\alpha$ emission detected at $S/N > 3$
\item $^c$Fraction of green valley galaxies classified as star forming ($f_{\mr{SF}}$), composite ($f_{\mr{comp}}$), or AGN ($f_{\mr{AGN}}$) based on the $[\mr N\mr{II}]/\mr H_\alpha$ vs. $[\mr O\mr{III}]/\mr H_\beta$ \citet{BPT} emission line diagram, using the \citet{Kewley01} extreme starburst classification line and the \citet{Kauffmann03a} pure star formation line
\item $^d$ Fraction of galaxies with low signal-to-noise ($S/N< 3$) in at least one of these emission lines, not included in the emission line classification
\end{tablenotes}
\end{threeparttable}
\label{tab:BPT}
\end{table}%

\section{Clustering Analysis}
\subsection{Projected Correlation Functions}
To separate spatial clustering from redshift space distortions, we 
first measure the correlation functions in radial direction $\pi$ and perpendicular direction $r_{\mr p}$ and then project out redshift space distortions. Specifically, we measure the (cross-)correlation function of galaxy samples $D_{X,Y}$ using the \citet{LS} estimator and its generalization for cross-correlation functions \citep{Szapudi98}
\be
\xi_{XY}(r_{\mr p}, \pi) = \left[\frac{D_X D_Y - D_X R_Y - D_Y R_X + R_X R_Y}{R_X R_Y}\right]\left(r_{\mr p}, \pi\right) ,
\ee
on a two-dimensional grid. Here $R_{X,Y}$ are associated random catalogs, $D D \left(r_{\mr p}, \pi\right)$, $D R \left(r_{\mr p}, \pi\right)$ and $R R \left(r_{\mr p}, \pi\right)$ are the (normalized) number of data-data, data-random, and random-random pairs at separation $\left(r_{\mr p}, \pi\right)$. We adopt linear binning in the radial component, logarithmic bins in perpendicular distance and measure the projected (cross-)correlation function as
\be
w_{XY}(r_{\mr p}) = 2\int_0^{\pi_{\mr{max}}} d\, \pi\, \xi_{XY}(r_{\mr p}, \pi)\,;
\ee
with $\pi_{\mr{max}} = 50\,\mr{Mpc}/h$.

\subsection{Measurements}
\label{sec:cor}
We generate random catalogs with the SDSS angular selection function and the angular selection function of the GALEX-SDSS cross-match catalog. As we have constructed volume-limited galaxy samples, and their color selected subsamples, with narrow redshift ranges allowing us to ignore redshift evolution effects, the random catalogs have uniform comoving density and do not need to account for the radial selection function. The random catalogs are oversampled compared to the galaxy catalogs by a factor $25$ for SDSS samples, and by a factor $100$ for the sparser $(NUV-r)$ selected samples. Increasing the the oversampling rate by a factor of two has no significant impact, indicating that the correlation function estimates have converged.

\begin{figure}
\includegraphics[width = 0.5\textwidth]{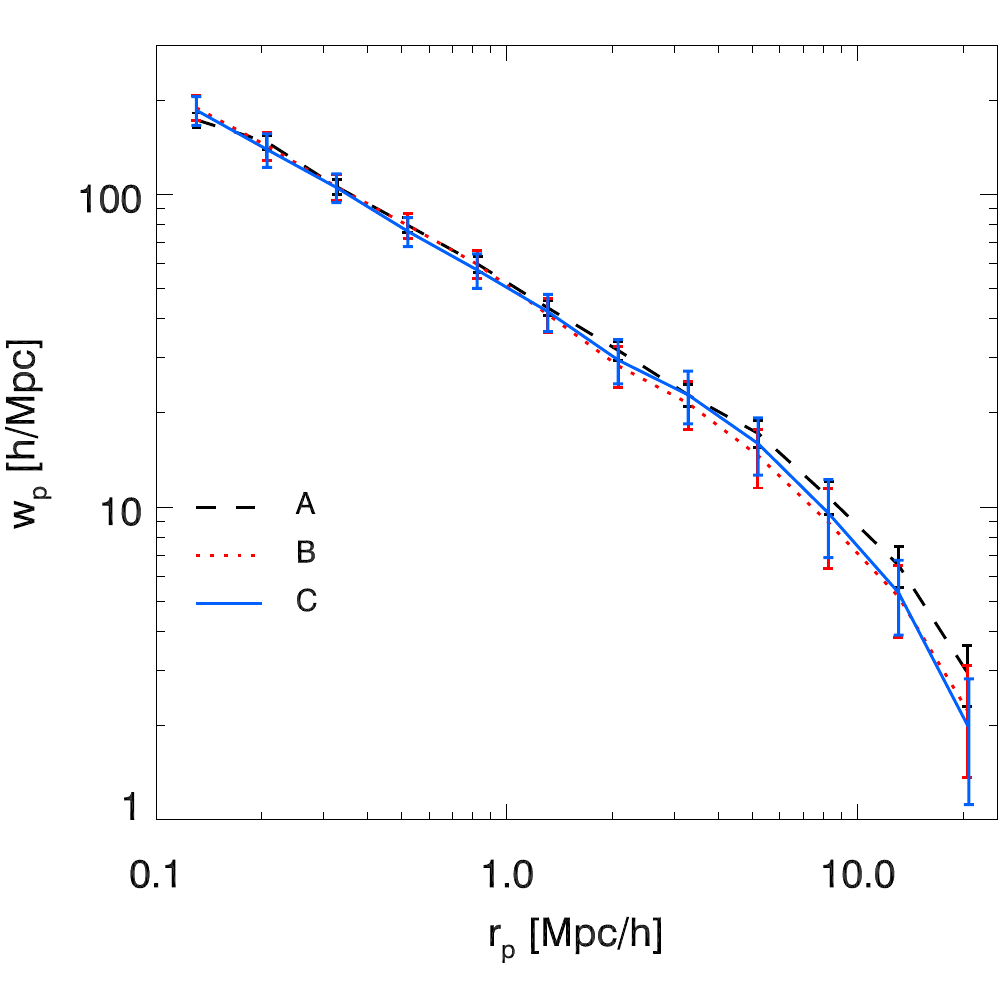}
\caption[Test of survey geometry effects on measured correlation functions]{Test of survey geometry effects on measured correlation functions. Different lines show the projected cross-correlation function between galaxies in the full SDSS footprint in the magnitude bin $[-19.5,-20]$ and with $(g-r) <0.8$ (A) with the same sample, (B) with the sample restricted to the combined survey area, (C) with GALEX detected galaxies in the same magnitude and color bin.}
\label{fig:wp_test}
\end{figure}

Figure~\ref{fig:wp_test} demonstrates that we have characterized the combined survey geometry sufficiently well to measure correlation functions in this patchy survey geometry. Here we show the correlation function between a galaxy sample in the full SDSS footprint in the magnitude bin $[-19.5,-20]$ and blue color ($(g-r) <0.8$) with different subsets of itself: The dashed line shows its auto correlation function. Next we consider the cross-correlation between this sample and its restriction to the footprint of the SDSS + GALEX combined catalog, which is shown by the dotted line. Compared to the full auto correlation function, this cross-correlation function may be affected by boundary effects associated with the correlation function estimator or finite volume effects, as we have reduced the volume probed by of one copy of the galaxy catalog by a factor of four. Note that in this case the angular selection function in the combined survey area is still given by the SDSS angular selection function. Next we further restrict one copy of the galaxy catalog to galaxies with $NUV$ detections, shown by the solid line. As the galaxy sample consists only of blue galaxies, these should all have $NUV$ detections, and any significant differences between the dotted and solid line would indicate a mis-characterization of the combined angular selection function. One copy of the galaxy catalog stays the same throughout the process, so that we measure the cross-correlation between samples with different footprints, which leads to better statistics and smaller finite volume effects than restricting the SDSS data to the combined footprint region as well.

\begin{figure*}
\includegraphics[width = 0.9\textwidth]{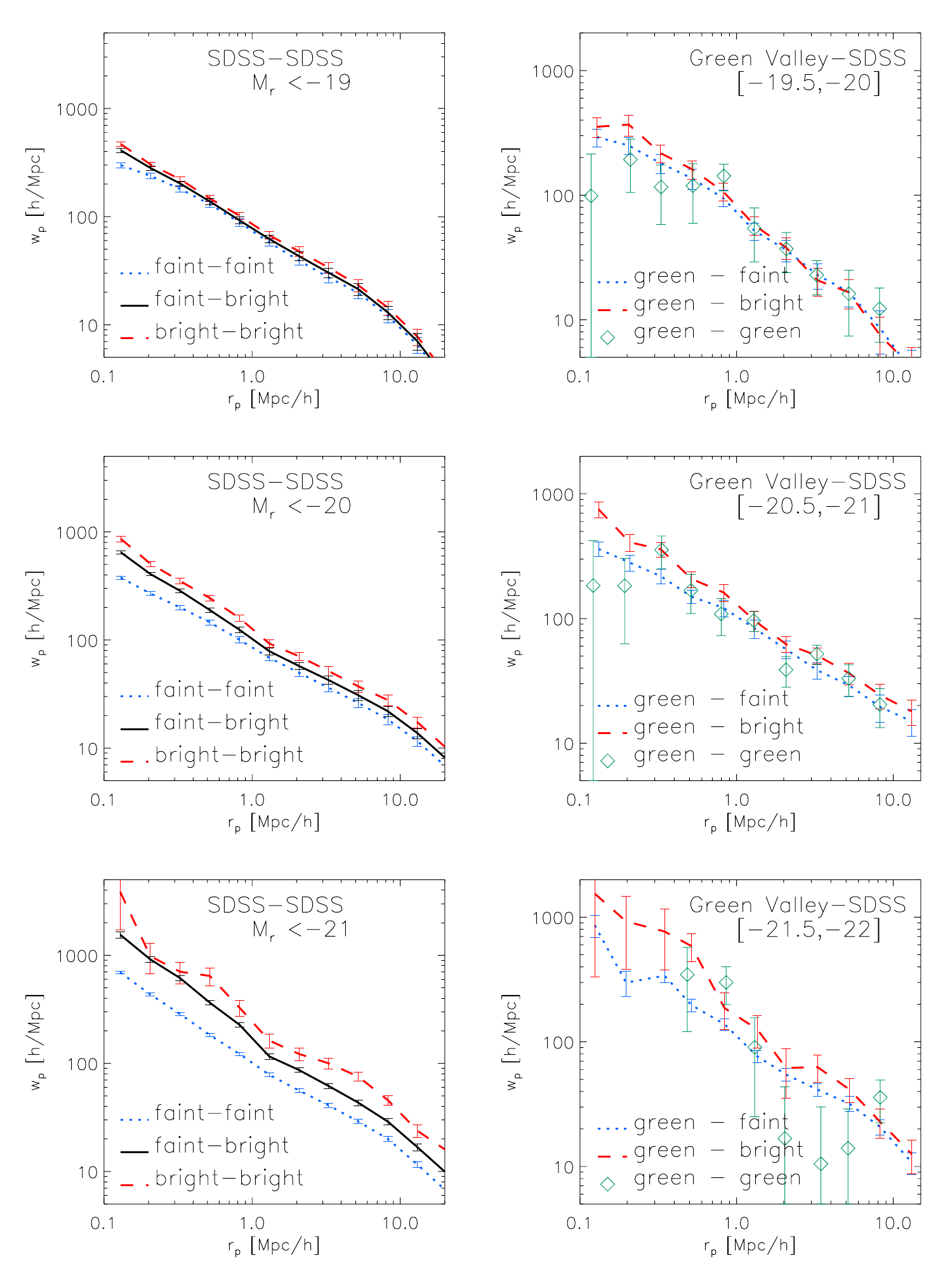}
\caption{Examples of measured cross-correlation functions. For comparison, we also show measurements the green valley galaxy auto correlation function.}
\label{fig:wp}
\end{figure*}

As described in detail in \citet{Zehavi10}, the clustering of the faintest SDSS luminosity threshold samples is subject to substantial sample variance effects due to the small volume probed by these low-redshift samples. As we are interested in a sparse subpopulation of these samples and are furthermore restricted to one fourth of the SDSS footprint area, these sampling effects are even more severe in our analysis. After reproducing the sub-volume tests of \citet{Zehavi10}, we find that the magnitude bin $[-19.5,-20]$ is the smallest sample for which we can obtain robust correlation function measurements. Examples of measured auto- and cross-correlation functions for SDSS galaxy samples and green valley galaxies are shown in Fig.~\ref{fig:wp}. For comparison, we also show measurements the green valley galaxy auto correlation function, for which we used random catalogs with an oversampling factor of 1000.

\begin{figure*}
\includegraphics[width = 0.45\textwidth]{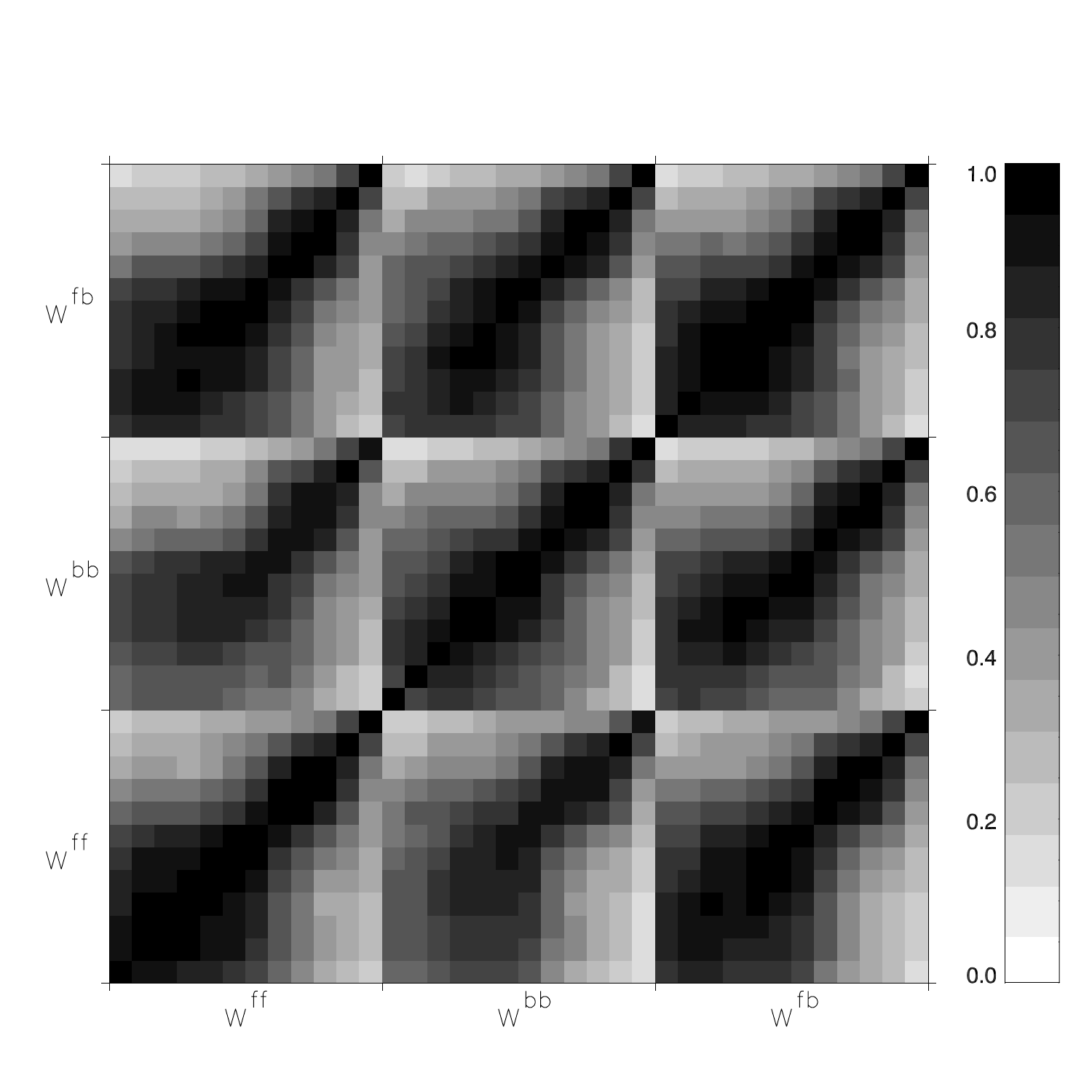}
\includegraphics[width = 0.45\textwidth]{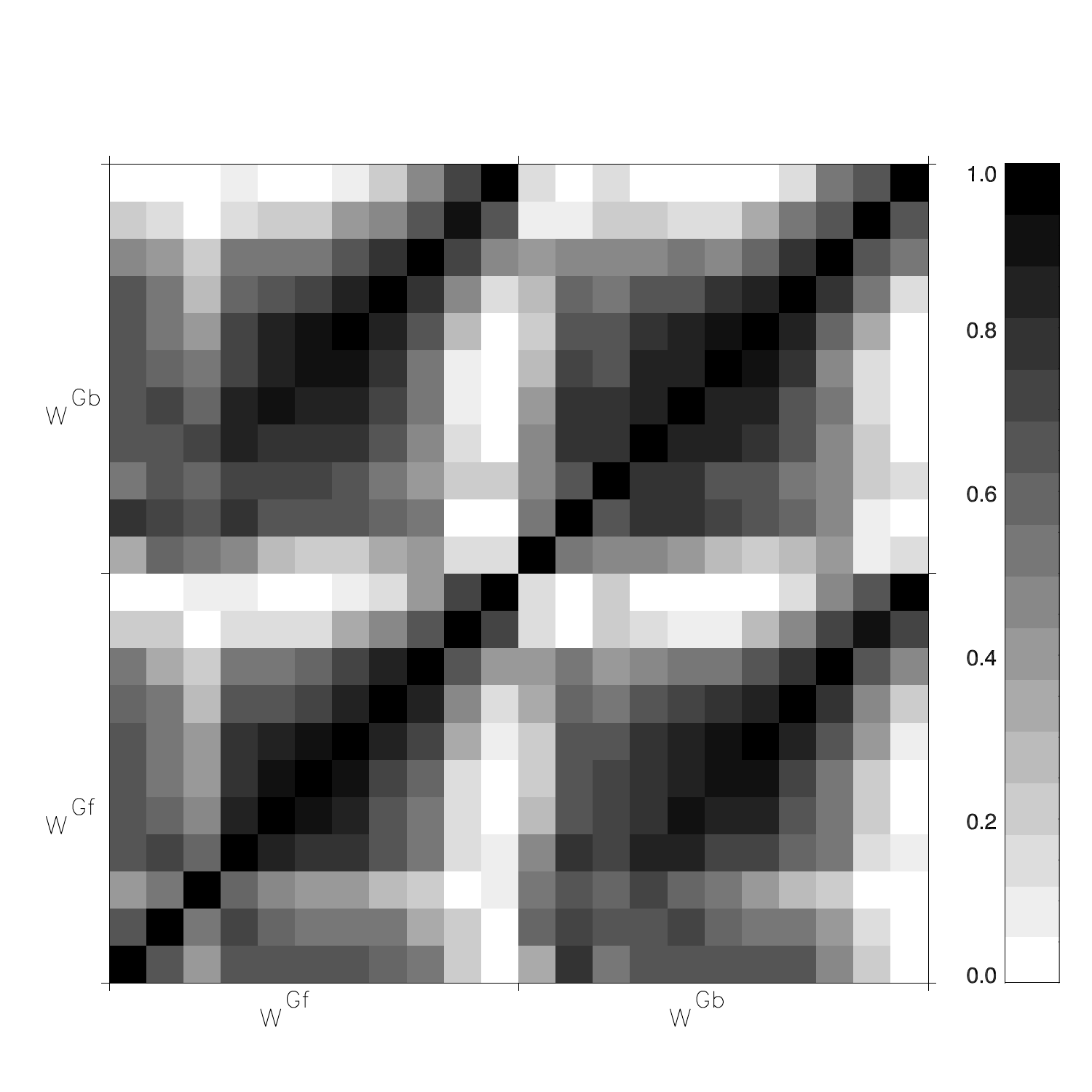}
\caption[Sample covariances]{Sample covariances. Top: Covariance between the different auto- and cross-correlation functions of the SDSS faint and bright sample associated with the magnitude bin $[-19.5,-20]$.
Bottom: Covariance of the cross-correlation function between the $[-19.5,-20]$ green valley sample and the corresponding SDSS faint and bright samples.\newline
In each block of these covariances perpendicular scales increase from left to right and bottom to top.}
\label{fig:cov}
\end{figure*}

We estimate the covariance of our correlation function measurements using bootstrapping with ``oversampling of subvolumes'' \citep{Norberg09} with an oversampling factor of 3, where number of subvolumes chosen with replacement $N_{\mr r}$ is equal to three times the number of subvolumes the data set is divided up into, $N_{\mr{sub}}$. \citet{Norberg09} find that this method gives robust error estimates that are in agreement with external estimates from mock catalogs. For correlation functions between two SDSS galaxy samples, we divide the SDSS footprint into 150 subsets of equal area. For correlation functions between one SDSS galaxy sample and one sample restricted to the combined footprint area, the division into equal area subsets is not clearly defined, and we choose subsets which contain equal number of random-random pairs at angular separation of $2^\circ$ in order to evenly sample the cross-correlation function on scales of a few $\mr{Mpc}/h$. Due to the smaller effective area of this restricted geometry, we only have 50 such subareas. Examples for both types of covariances are shown in Fig.~\ref{fig:cov}. As noted by \citet{Hartlap07}, estimated covariances are a biased estimate of the inverse covariance with the bias depending on the number of data points,$p$, and the number of independent data sets, $n$.
If the mean is estimated from the data, an unbiased estimate of the inverse covariance is given by
\be
\hat{\mathbf C}_{\mr{unbiased}}^{-1} 
=\frac{n-p-2}{n-1}\hat{\mathbf C}^{-1}
 \approx\left(1-\frac{p}{n}\right)\hat{\mathbf C}^{-1}
\,.
\label{eq:Cinv}
\ee
As bootstrap realizations are not independent, we cannot apply Eq.~\ref{eq:Cinv} directly with $n = N_{\mr r}$. Instead, we assume
\be
\hat{\mathbf C}_{\mr{unbiased}}^{-1} 
 \approx\left(1-m \frac{p}{N_{\mr r}}\right)\hat{\mathbf C}^{-1}\,,
\ee
and follow the calibration method described in \citet{Eifler08}: We measure $\mr{tr}(\hat{\mathbf C}^{-1})$ repeatedly varying $N_{\mr r}$ with constant binning and oversampling rate, and determine $m$ as the slope of $1/\mr{tr}(\hat{\mathbf C}^{-1})$ with in $p/N_{\mr r}$. Specifically, we varied $N_\mr r$ using $N_{\mr{sub}}=$(120,135,150,165,180) for the SDSS footprint, and $N_{\mr{sub}}=$(40,45,50,55,60) for the GALEX-SDSS footprint.

We were unable to obtain stable, invertible covariances for the most luminous green valley galaxy sample. Hence we restrict our analysis of this sample to large scales (section \ref{sec:blin}) where it was possible to measure converged and invertible covariances.
\subsection{Results: Large-Scale Bias}
\label{sec:blin}
\begin{figure*}
\includegraphics[width = \textwidth]{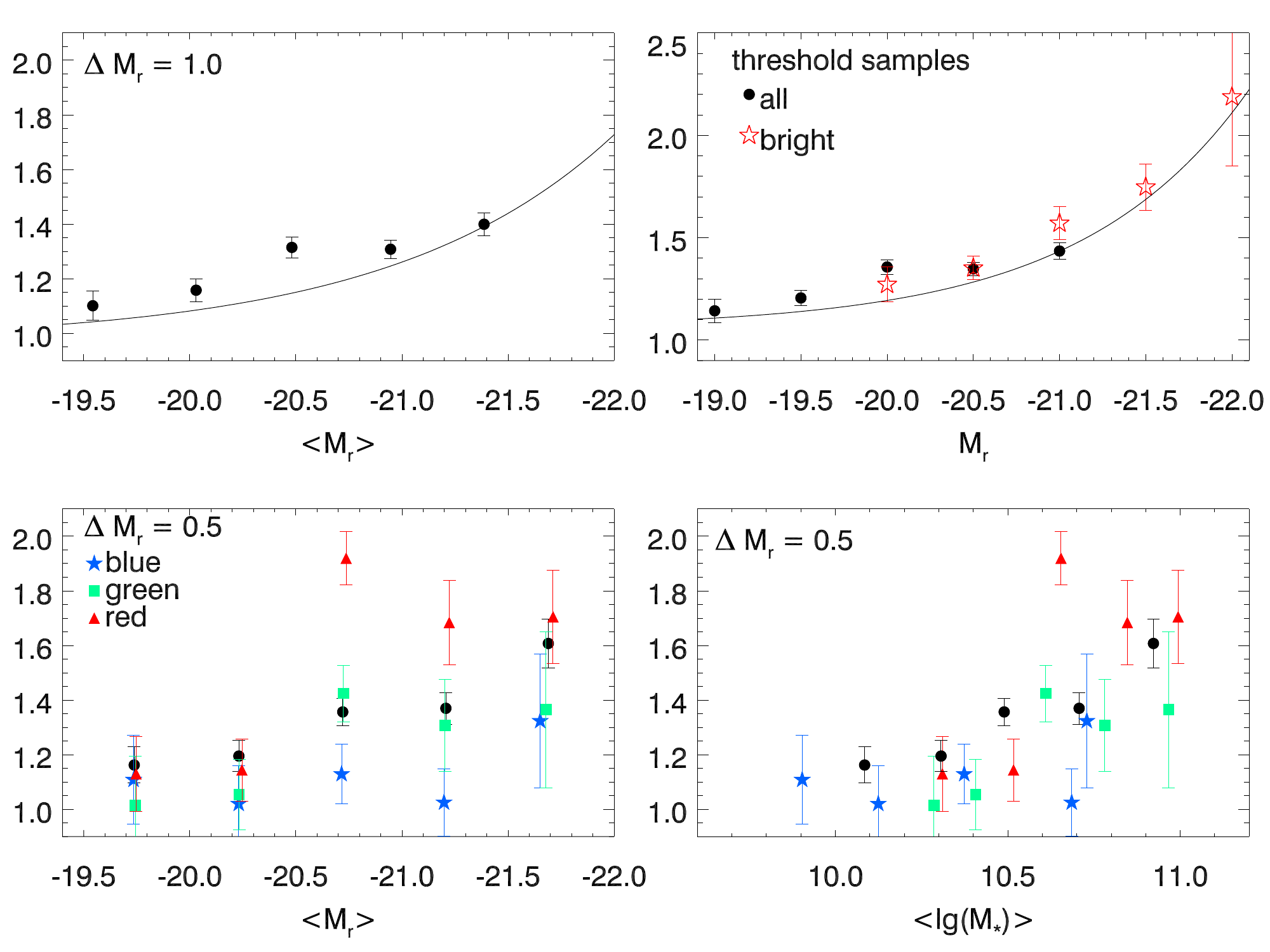}
\caption[Linear galaxy bias measurements from large-scale correlation function]{Linear galaxy bias measurements obtained from fits to the large-scale correlation function. Top: Linear bias as a function of luminosity for different luminosity bin samples with bin width $\Delta M_r = 1.0$ (left), and threshold samples (right). The lines show best-fit relations from \citet{Zehavi10}.\newline
Bottom: Linear bias as a function of $(NUV-r)$ color and luminosity (left) or stellar mass (right), for galaxy samples with luminosity bin width $\Delta M_r = 0.5$.} 
\label{fig:ls_bias}
\end{figure*}

Based on the correlation function measurements described in the previous section, we can measure the large-scale galaxy bias by fitting the projected correlation functions with theoretical matter correlation functions times a linear bias factor. Specifically, we fit measured correlation functions over the range $3-25$ Mpc$/h$ to the theoretical predictions for the projected matter correlation function, including the full data covariance. Figure~\ref{fig:ls_bias} shows the resulting luminosity bias relation.
The top two plots are for binned and threshold samples of SDSS galaxies, and the lines are fits from the analysis of galaxy clustering in SDSS DR7 by \citet{Zehavi10}. Overall, we find good agreement with their results. The $M_{\mr r}< -20$ galaxy threshold sample and it subsamples deviate from the best-fit bias relation. As detailed in Tab.~\ref{tab:sample1}, these samples are centered around the redshift of the Sloan Great Wall, which leads to excess clustering in this and neighboring samples.\footnote{This was also noted by \citet{Zehavi10} who exclude the redshift range of the Sloan Great Wall from their analysis of luminosity bin galaxy samples} This effect is enhanced in the lower plots, which show bias as a function of $(NUV-r)$ color and luminosity or mean stellar mass. Here the clustering of red galaxies is strongly enhanced in the Sloan Great Wall.
\section{Halo-Occupation Distribution Modeling}
At the level of individual halos, a halo-occupation distribution (HOD) model \citep[e.g.,][]{BW02} describes the relation between galaxies and halo mass in terms of the probability $P(N,M_{\mr h})$ that a halo of given mass $M_{\mr h}$ contains $N$ galaxies. To describe the two-point clustering of galaxies, we need models for first and second moment of the HOD, $\ensav{N|M_{\mr{h}}}$ and $\ensav{N(N-1)|M_{\mr{h}}}$. Following \citet{Zheng05}, we separate galaxies into central and satellite galaxies. By definition, a halo contains either zero or one central galaxy, and it may host satellite galaxy only if it contains a central galaxy, which motivates the form
\be
 \ensav{N(M_{\mr h})} = \ensav{N_{\mr{c}}|M_{\mr h}}\left(1 + \ensav{N_{\mr{s}}|M_{\mr h}}\right),
 \label{eq:NM}
\ee
with $\ensav{N_{\mr{c/s}}|M_{\mr h}}$ the average number of central/satellite galaxies in a halo of mass $M_\mr{h}$.
\subsection{HOD Parameterization}
While the assumptions in a HOD model describing the properties of dark matter halos are generally agreed upon (see section \ref{sec:HODwp} for details), the form of the relation between galaxies and halos (equation (\ref{eq:NM})) is less well constrained and leaves more room for experiments. We motivate the details our implementation next.
\subsubsection{SDSS Samples}
\label{sec:HODsdss}
We base our model for SDSS galaxy samples on the HOD parameterization of \citet{Zehavi10} for luminosity thresholds samples with absolute $r$-band magnitude $M_r<M_r^t$,
\be
\ensav{N(M_{\mr h}|M_r^t)} = \frac{1}{2}\left[1+\mr{erf}\left(\frac{\log M_{\mr h}-\log M^{\mr t}_{\mr{min}}}{\sigma^{\mr t}_{\mr{log}M}}\right)\right]\left[1+\left(\frac{M_{\mr h}-M^{\mr t}_0}{M_1^{\prime{\mr t}}}\right)^{\alpha_{t}}\right]\,,
\label{eq:NM}
\ee
with model parameters $M^{\mr t}_{\mr{min}}, M^{\mr t}_0, M_1^{\prime{\mr t}}, \sigma^{\mr t}_{\mr{log}M}, \alpha_{t}$. The central galaxy occupation function is a softened step function with transition mass scale $M^{\mr t}_{\mr{min}}$, which is the halo mass in which the median central galaxy luminosity corresponds to the luminosity threshold, and softening parameter $\sigma^{\mr t}_{\mr{log}M}$ which is related to the scatter between galaxy luminosity and halo mass. The normalization of the satellite occupation function, $M_1^{\prime{\mr t}}$, and cut-off scale $M^{\mr t}_0$ are related to $M_1$, the mass scale at which a halo hosts at least on satellite galaxy ($N_{\mr s}(M_1) =1)$); finally $\alpha_{t}$ is the high-mass slope of the satellite occupation function. This parametrization was found to reproduce the clustering of SDSS and CFHTLS galaxies (Coupon et al. 2012) well over a large range of luminosity thresholds and redshifts.

The HOD model for a binned galaxy sample with $M_r^{t_2}< M_r < M_r^{t_1}$ is typically calculated from model fits to luminosity threshold samples as 
\be
\ensav{N(M_{\mr h}|M_r^{\mr t_1},M_r^{\mr t_2})}= \ensav{N(M_{\mr h}|M_r^{\mr t_1})}-\ensav{N(M_{\mr h}|M_r^{\mr t_2})}\,.
\label{eq:NMt}
\ee 
While we note that the results of \citet{Zehavi10} favor a somewhat steeper slope of the satellite distribution for the most luminous galaxy samples in our analysis, we set $\alpha = 1$ for all SDSS galaxy samples. This is in overall agreement with previous results for the luminosity range of interest, and makes differencing the HOD of neighboring samples numerically more stable.
Hence our model has 4 free parameters for a luminosity threshold sample, and 8 free parameters for a luminosity bin sample. Without further constraints, such a parameterization of luminosity bin samples has too many degrees of freedom for general applications. However, it has the advantage that the HODs of neighboring luminosity bins are consistent with each other, and we use this parameterization to fit the different correlation functions among our SDSS faint and bright samples, resulting in 8 parameters for the SDSS HODs in each volume-limited sample. 

Furthermore, we assume the radial distribution of all color independent galaxy samples to follow the dark matter distribution. This assumption is supported by the results of \citet{Watson12} who studied the small scale clustering of SDSS galaxies. While these authors found an enhanced clustering of luminous galaxies on small scales ($r_{\mr p} < 0.05 Mpc/h$) compared to an NFW distribution, their galaxy correlation function measurements agree very well with the predicted dark matter clustering over the radial scales and luminosity range of interest for our analysis.
\subsubsection{Luminosity and Color bin Samples}
\label{sec:HODnuv}
For a $(NUV-r)$ selected galaxy sample ($X$), which is measured in one narrow 0.5 mag bin per sample volume, we need a more compact description of the HOD and we model the central galaxy term as a clipped Gaussian,
\be
\ensav{N_{\mr c}(M_{\mr h}, X)} = \min(\frac{A_X}{\sigma_X}\sqrt{2\pi } \exp\left(\frac{-(\log M_\mr{h} - \log M_{\mr{c}}^X)^2}{2\sigma_X^2}\right),1)\,,
\ee
with free parameters $A_X$, $\sigma_X$ and $M_{\mr{c}}^X$. Here the clipping enforces that a halo does not have more than one central galaxy.

The auto correlation function of color selected galaxies by definition is only sensitive to galaxy pairs of the same color. Hence HOD models require assumptions on the relation between the colors of central and satellite galaxies, and in particular need to account for central galaxies which are not part of sample \citep[e.g.][]{Simon09, Skibba}. In contrast, modeling the cross correlation between a color selected galaxy sample and the full (color independent) galaxy population with the same luminosity threshold does not require such assumptions. This allows us to simply write the condition that a halo has to contain a central galaxy in order to host satellite galaxies in terms of central galaxy occupation function of the full (color independent) luminosity threshold sample with luminosity threshold $t_X$ equal to the minimum luminosity of the luminosity bin under consideration, 
\be
\ensav{N_{\mr s}(M_{\mr h}, X)}  = A_X  \frac{1}{2}\left[1+\mr{erf}\left(\frac{\log M_{\mr h}-\log M^{\mr t_X}_{\mr{min}}}{\sigma^{\mr t_X}_{\mr{log}M}}\right)\right] \left(\frac{M_{\mr h}}{M_1^X}\right)^{\alpha_X}\,
\ee
which is characterized by two free parameter, $M_1^X$ and $\alpha_X$.

Note that the correlation function of a binned sample is independent of the normalization parameter $A_X$, which is determined by the galaxy number density.

Motivated by observations finding red satellite galaxies to be radially more concentrated than blue galaxies \citep[e.g.][]{V10,Guo12}, we introduce another free parameter $f_X$ which describes the NFW concentration, $c_X$, of a color selected galaxy sample relative to that of dark matter, 
\be
c_X(M_{\mr h}) = f_X c(M_{\mr h})\,.
\ee
\subsection{Relation to Correlation Functions}
\label{sec:HODwp}
The halo model prediction for the real-space correlation function takes the form
\be
1 + \xi(r) = \left(1+\xi^{1\mr h}\left(r\right)\right) + \left(1 +\xi^{2\mr h}\left(r\right)\right)\,,
\ee
where $(1+\xi^{1\mr h})$ is proportional to the number of galaxy pairs residing in the same halo (\emph{one-halo term}), and the \emph{two-halo term} $(1 +\xi^{2\mr h})$ is proportional to the number of galaxy pairs occupying different halos. The model real-space correlation function is related to the projected correlation function as
\be
w(r_{\mr p}) = 2\int_0^{\pi_{\mr{max}}} d\pi\, \xi\left(\sqrt{r_{\mr p}^2 + \pi^2}\right)\,.
\ee
We will now describe the computation of these terms in detail. In order to evaluate these expressions numerically, we define halos to enclose a spherical overdensity of 200 times the mean background density and assume that their density distribution follows a NWF profile \citep{NFW97} with the halo mass--concentration relation of \citet{Bhattacharya11}; furthermore we use the fitting functions of \citet{Tinkerm} and \citet{Tinkerb} for the halo mass function and halo bias relation. Unless stated otherwise, we assume that the galaxy distribution follows the halo density profile.
\subsubsection{One-Halo Term}
We split the computation of the one-halo term into then clustering of central and satellite galaxy $\xi^{1,c-s}$ and satellite-satellite clustering $\xi^{1,s-s}$ within the same halo. The central-satellite term is given by
\beq
\nonumber 1+\xi^{1,c-s}_{XY}(r) \!\!\!&=&\!\!\! \!\!\!\frac{1}{\bar n_X \bar n_Y}
\int_{M_{\mr{vir}(r)}}^\infty\!\!\!\!\!\!  dM_{\mr h}\,\frac{d n}{d M_{\mr h}}
\bigg(
\ensav{N_c(M_{\mr h},X) N_s(M_{\mr h},Y)}
\rho_Y(r|M_{\mr h})\\
&&+ \ensav{N_c(M_{\mr h,}Y) N_s(M_{\mr h},X)}\rho_X(r|M_{\mr h})\bigg)\,,
\label{eq:xics}
\eeq
where $dn/dM_{\mr h}$ denotes the halo mass function, with $\rho_X (r|M_{\mr h})$ the normalized radial distribution of galaxy population $X$ within the halo, and with
\be
\bar n_X = \int_0^\infty d M_{\mr h}\,\frac{dn}{dM_{\mr{h}}}\ensav{N(M_{\mr h}|X)}\,.
\ee
The term $\ensav{N_c(M_{\mr h},X) N_s(M_{\mr h},Y)}$ in equation (\ref{eq:xics}) is equal to the average number galaxy pairs with a central galaxy from sample $X$ and a satellite galaxy from sample $Y$ in a halo of mass $M_{\mr h}$. From the definition of satellite galaxy this term evaluates to $\ensav{N_c(M_{\mr h},M_r^\mr{t}) N_s(M_{\mr h},M_r^\mr{t})}= \ensav{N_s|M_{\mr h},M_r^\mr{t})}$ for the auto correlation of a luminosity threshold sample \citep{Zheng05}. However, when considering binned samples or cross-correlations between different samples, the central galaxy of a halo hosting satellite galaxies from the sample $Y$ need not be from sample $X$, and we use $\ensav{N_c(M_{\mr h},X) N_s(M_{\mr h},Y)} =\ensav{N_c|M_{\mr h},X}\ensav{N_s|M_{\mr h},Y}$ \citet{Miyaji11}.\\
If samples $X$ and $Y$ are disjunct, the satellite-satellite term is given by
\beq
\nonumber 1+\xi^{1,s-s}_{XY}(r) &=& \frac{1}{\bar n_X \bar n_Y}
\int_{M_{\mr{vir}(r)}}^\infty\!\!\!\!\!\!  dM_{\mr h}\frac{d n}{d M_{\mr h}} 
\ensav{N_s(M_{\mr h},X) N_s(M_{\mr h},Y)}\\
&&\;\;\;\;\;\;\;\;\;\;\;\;\times \left(\rho_X\ast\rho_Y\right)(r|M_{\mr h})\,,
\label{eq:xiss}
\eeq
where $\left(\rho_X\ast\rho_Y\right)(r|M_{\mr h})$ denotes the convolution of radial galaxy distributions $\rho_X$ and $\rho_Y$, and where the average number of satellite pairs is given by $\ensav{N_s(M_{\mr h},X) N_s(M_{\mr h},Y)} = \ensav{ N_{\mr s}|M_{\mr h},X}\ensav{ N_{\mr s}|M_{\mr h},Y}$.\\
To model auto correlations function, the number of galaxy pairs is modified to
\beq 
\nonumber 1+\xi^{1,s-s}_{XX}(r) &=& \frac{2}{\bar n_X \bar n_X}
\int_{M_{\mr{vir}(r)}}^\infty\!\!\!\!\!\!  dM_{\mr h}\frac{d n}{d M_{\mr h}} 
\frac{\ensav{N_s(M_{\mr h},X) (N_s(M_{\mr h},X)-1)}}{2} \\
&&\;\;\;\;\;\;\;\;\;\;\;\;\times \left(\rho_X\ast\rho_X\right)(r|M_{\mr h})\,.
\eeq
Assuming that satellite galaxies are Poisson distributed, the number of pairs evaluates to $\ensav{N_{\mr s}(N_{\mr s}-1)} = \ensav{N_{\mr s}}^2$.\\
\subsubsection{Two-Halo Term}
On scales above $\sim\!5 \mr{Mpc}/h$, the clustering of galaxies follows the large-scale clustering of dark matter halos, and it is modeled as function of the dark matter correlation function $\xi^{\mr{mm}}$,
\be
\xi^{2\mr h}_{XY}(r) \approx b_X b_Y \xi^{\mr{mm}}(r)\,.
\ee
Here $b_X$ denotes the bias parameter of galaxy sample $X$, which we calculate as
\be
b_{X} = \frac{1}{\bar n_X}\int_0^\infty d M_{\mr h}\,\frac{dn}{dM_{\mr h}} b_{\mr h}(M_{\mr h})\ensav{N(M_{\mr h}|X)} \,,
\ee
where $b_{\mr h}$ is the halo bias parameter.

On intermediate scales one needs to account for the distribution of galaxies within different halos and halo exclusion, i.e., the fact that two halos contribution to the two-halo term do not overlap. Following the spherical halo exclusion model of \citet{Tinker05}, we restrict the calculation of the two-halo term at separation $r$ to halos with $R_{\mr{vir},1} + R_{\mr{vir},2} \le r$. The effect of the distribution of galaxies within the different halos on the correlation function is given by the convolution of their respective density profiles. As this requires convolving many different halo profiles, we calculate the two-halo term is calculated in Fourier space:
\beq
\nonumber P_{XY}^{2\mr h} (k,r) &=& P_m(k)\frac{1}{\bar n_X^\prime \bar n_Y^\prime(r)}\\
\nonumber&&\!\!\!\!\times \int_{M_{\mr{min}}}^{M_{\mr{lim,1}}(r)} \!\!\!\!d M_1\, \frac{d n}{dM_1} \ensav{N|M_1,X}b_{\mr h}(M_1)\tilde\rho_X(k,M_1)\\
&&\!\!\!\!\times \int_{M_{\mr{min}}}^{M_{\mr{lim,2}}(M_1,r)} \!\!\!\!\!\!\!\!\!\!\!d M_2\, \frac{dn}{dM_2} \ensav{N|M_2,Y}b_{\mr h}(M_2)\tilde\rho_Y^\ast(k,M_2)\,,
\label{eq:Pgg}
\eeq
where $M_{\mr{lim},1}$ is the maximum halo mass such that $R_{\mr{vir}}(M_{\mr{lim},1}) = r-R_{\mr{vir}}(M_{\mr{min}})$ with $M_{\mr{min}}$ the minimum halo mass of the HOD,  where $M_{\mr{lim},2}$ is defined by $R_{\mr{vir}}(M_{\mr{lim},2}) = r-R_{\mr{vir}}(M_{\mr{lim},1})$, and where $\tilde \rho_X$ denotes the Fourier transform of the normalized galaxy distribution $\rho_X$. $\bar n_X^\prime \bar n_Y^\prime(r)$ denotes the number density of galaxy pairs restricted to non-overlapping halos at separation $r$
\be
\bar n_X^\prime \bar n_Y^\prime(r) = 
\int_{M_{\mr{min}}}^{M_{\mr{lim,1}}(r)} \!\!\!\! \!\!\!\!d M_1\, \frac{d n}{dM_1} \ensav{N|M_1,X}
\int_{M_{\mr{min}}}^{M_{\mr{lim,2}}(M_1,r)} \!\!\!\!\!\!\!\!\!\!\!d M_2\, \frac{dn}{dM_2} \ensav{N|M_2,Y}\,.
\ee
The two-halo correlation function is obtained from the power spectrum by
\be
\xi_{XY}^{2\mr h\prime}(r) = \frac{1}{2\pi^2}
\int_0^\infty d k\, k^2 
\frac{\sin(kr)}{kr}
P_{XY}^{2\mr h}(k,r)\,.
\ee
As $\xi_{XY}^{2\mr h\prime}(r)$ has been obtained from a (radius -) restricted sample of galaxy pairs, it is converted to a probability for the whole sample by
\be
1+\xi_{XY}^{2\mr h}(r) = \frac{\bar n_X^\prime \bar n_Y^\prime(r)}{\bar n_X \bar n_Y}\left(1+\xi_{XY}^{2\mr h\prime}(r)\right)\,.
\ee
\subsection{Analysis}
As described in section  \ref{sec:cor}, for each luminosity bin sample of interest we measure the projected auto and cross-correlation functions of the SDSS faint and bright galaxy samples
\be
\left(\mathbf w_{\mr{ ff}}, \mathbf w_{\mr{fb}}, \mathbf w_{\mr{bb}}\right) \equiv \mathbf W_{\mr{S}}\,,
\ee
where we have introduced the correlation function data vector  $\mathbf w=(w(r_{\mr p,1}),w(r_{\mr p,2}),...,w(r_{\mr p,N_{\mr bin}}))$ , and the cross-correlation between $(NUV-r)$ color selected luminosity bin samples and the two SDSS galaxy samples  
\be
\left(\mathbf w_{X\mr f}, \mathbf w_{X\mr b} \right)\equiv \mathbf W_{X}\,,
\ee with $X\in\{\mr{blue,\, green,\, red}\}$).

Ideally one would fit all these cross-correlation functions simultaneously, however this is not practicable: As the $(NUV-r)$ selected galaxy samples are restricted to GALEX + SDSS overlap area, obtaining a joint covariance for the SDSS reference samples and the color selected sample ($\mr{Cov}(\mathbf w_{\mr{ ff}}, \mathbf w_{\mr{fb}}, \mathbf w_{\mr{bb}},\mathbf w_{X\mr f}, \mathbf w_{X\mr b})$) would require restricting the SDSS clustering analysis to the combined SDSS + GALEX footprint, which would discard 75\% of the SDSS area.\footnote{Also note that even if one was willing to discard most of the SDSS data, obtaining an invertible joint covariance for the five different correlation functions, sampled with $N_{\mr{bin}}$ radial bins, would require dividing the joint footprint into more than $5\times N_{\mr{bin}}$ equal-area jack knife regions $N_{\mr{sub}}$. Additionally, the correction factor required to obtain an unbiased estimate of the inverse covariance scales as the ratio of the number of bins (data vector variables) to the number of data sets \citep{Hartlap07}, resulting either in very large error bars ($N_{\mr{sub}}\sim 5 N_{\mr{bin}}$) or restricting the analysis to very small scales ($N_{\mr{sub}}\gg 5 N_{\mr{bin}}$).}

Instead, we first model the SDSS correlation functions and galaxy number densities with an eight parameter HOD described in section \ref{sec:HODsdss}, and then fit the color bin sample HOD (section \ref{sec:HODnuv}) using the model for the SDSS samples obtained in the previous step; using the full (non block diagonal) data covariances (Fig.~\ref{fig:cov}) in each step. This method assumes that the color sample - SDSS sample cross-correlations ($\mathbf w_{X\mr f}, \mathbf w_{X\mr b}$) contain little information on the HOD of the SDSS sample compared to the SDSS internal correlation functions used in the first step of the fitting procedure. This assumption is well motivated by statistical uncertainties as the color selected samples are over an order of magnitude smaller than the SDSS reference samples. We propagate correlated uncertainties in the HOD model parameters for the SDSS reference sample to the HOD of the color bin sample by marginalizing over 15 randomly chosen models for the SDSS HOD.

Specifically, we compute the $\chi^2$ as
\beq
\nonumber \chi^2 &=& \left(\mathbf W_Y^{\mr{data}}-\mathbf W_Y^{\mr{model}}\right)\mr{Cov}^{-1}(\mathbf W_Y) \left(\mathbf W_Y^{\mr{data}} -\mathbf W_Y^{\mr{model}}\right)\\
&&+ \left(\mathbf n_Y^{\mr{data}} - n_Y^{\mr{model}}\right)\mr{Cov}^{-1}(\mathbf n_Y)\left(\mathbf n_Y^{\mr{data}} - n_Y^{\mr{model}}\right)\,,
\eeq
where $Y \in\{S,X\}$, with galaxy number densities $\mathbf n_S = (n_{\mr f}, n_{\mr b})$ or $\mathbf n_X = n_X$, and with the statistical error on the number densities $\mr{Cov}(\mathbf n_Y)$ estimated from field to field variations. The HOD parameter space is explored using a Markov Chain Monte Carlo method with a multi-variate Gaussian proposal function and flat priors $\{\log_{10} M_{\mr min},\log_{10} M_{\mr c}^X,\log_{10}M_1,\log_{10}M_1^\prime \} \in[11,17]$, $\{\sigma_{\mr{log}M}, \sigma_X\}\in[0.05,1.0]$, $\{\alpha_X,f_X\}\in[0.5,2.0]$,and $\log_{10}M_0\in[8,15]$. 
At each step a new set of HOD parameters is always accepted if $\chi^2_{\mr{new}} \leq \chi^2_{\mr{old}}$, and it is accepted with probability $\exp(-(\chi^2_{\mr{new}}-\chi^2_{\mr{old}})/2)$ if $\chi^2_{\mr{new}} > \chi^2_{\mr{old}}$. The typical chain length is 20000, and we compare 10 chains of length 20000 and one chain of length 100000 to test convergence. 
\subsection{Results}
Our best-fit HOD model parameters for the SDSS samples and their marginalized $1\sigma$ errors are given in Tab.~\ref{tab:HODsdss}. Our results agree well with the corresponding luminosity threshold samples in the analysis of \citet{Zehavi10}, and we confirm the overall trend of characteristic halo masses for hosting central and satellite galaxies with luminosity threshold. For a detailed comparison note that these two analyses use different fitting formulae for the halo mass function, halo bias and halo mass -- concentration relations.

\begin{figure*}
\includegraphics[width = 0.9\textwidth]{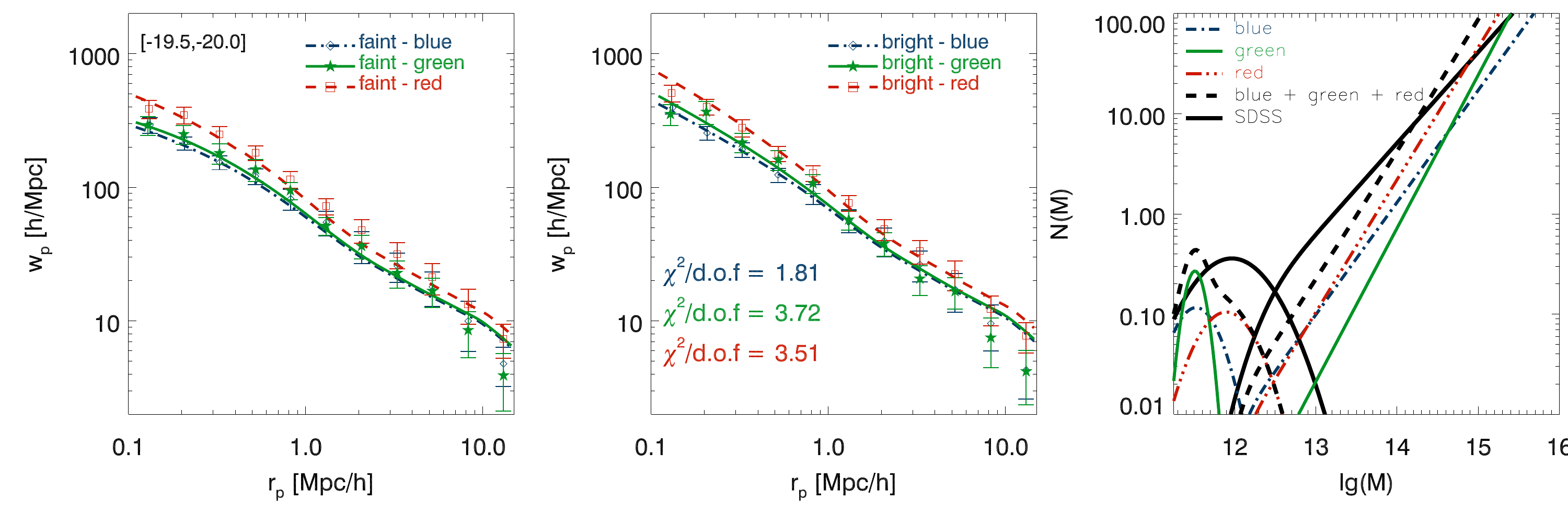}
\includegraphics[width = 0.9\textwidth]{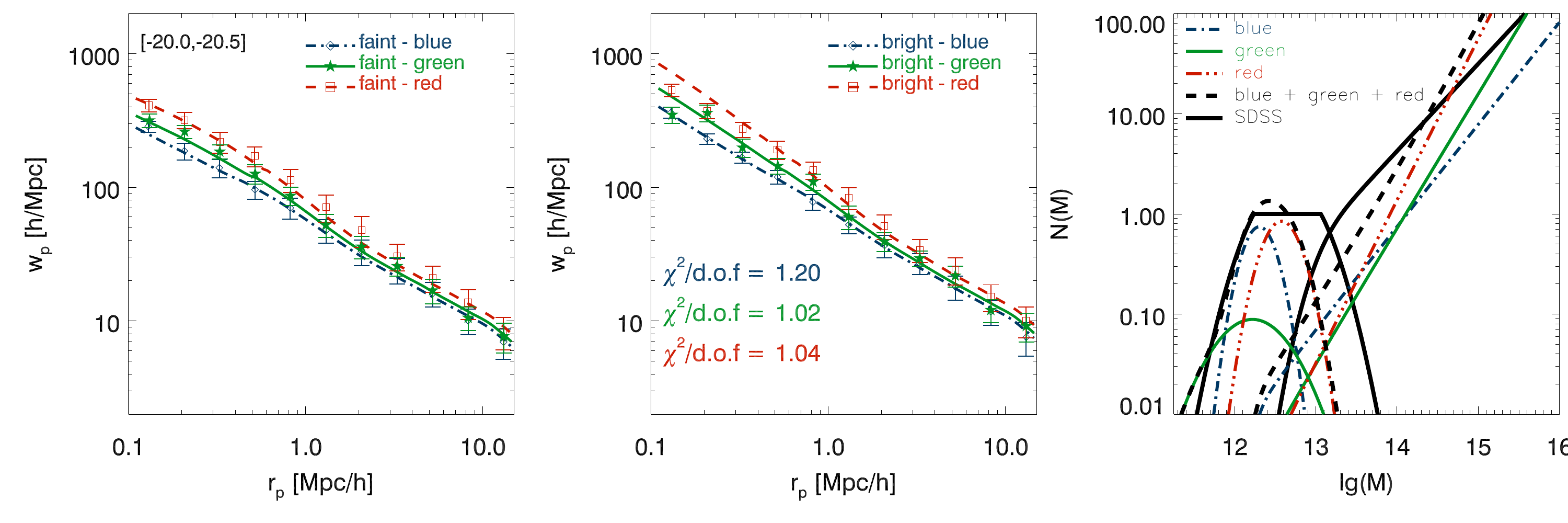}
\includegraphics[width = 0.9\textwidth]{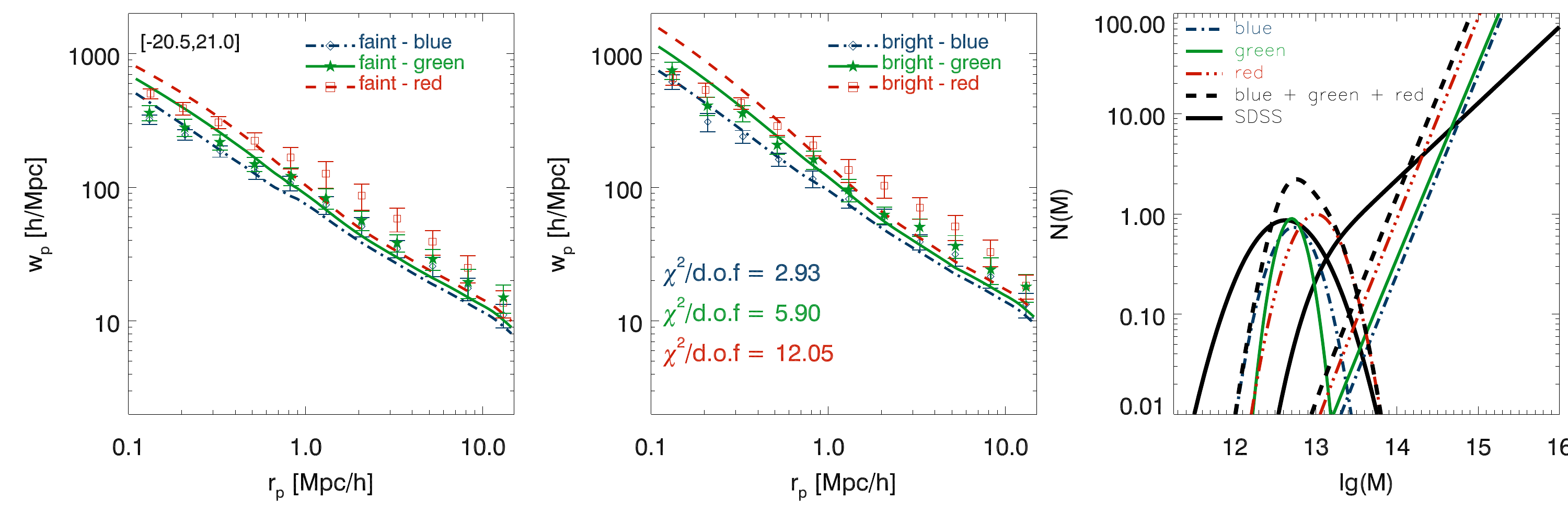}
\includegraphics[width = 0.9\textwidth]{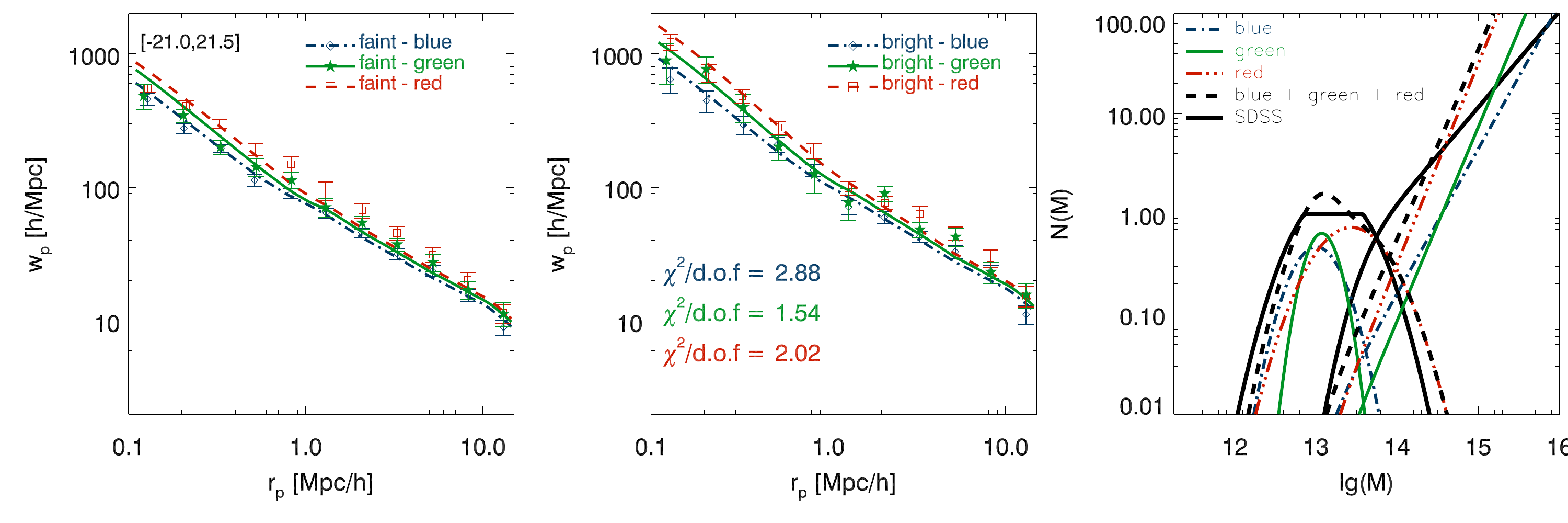}
\caption[Best-fit HOD for (NUV-r) selected samples]{Each row shows the measured correlation functions and best-fit HOD of $(NUV-r)$ selected galaxy samples for one luminosity bin. The left/middle panel show the cross-correlation measurements using the faint/bright sample and their joint fit. We list the reduced $\chi^2$ of these fits in the middle panel. The right panel shows the color sample HOD derived from fitting these cross-correlation functions, the sum of all the color samples, and the best-fit HOD of all SDSS galaxies in the same luminosity bin.}
\label{fig:hodfit}
\end{figure*}

\begin{figure}
\includegraphics[width = 0.5 \textwidth]{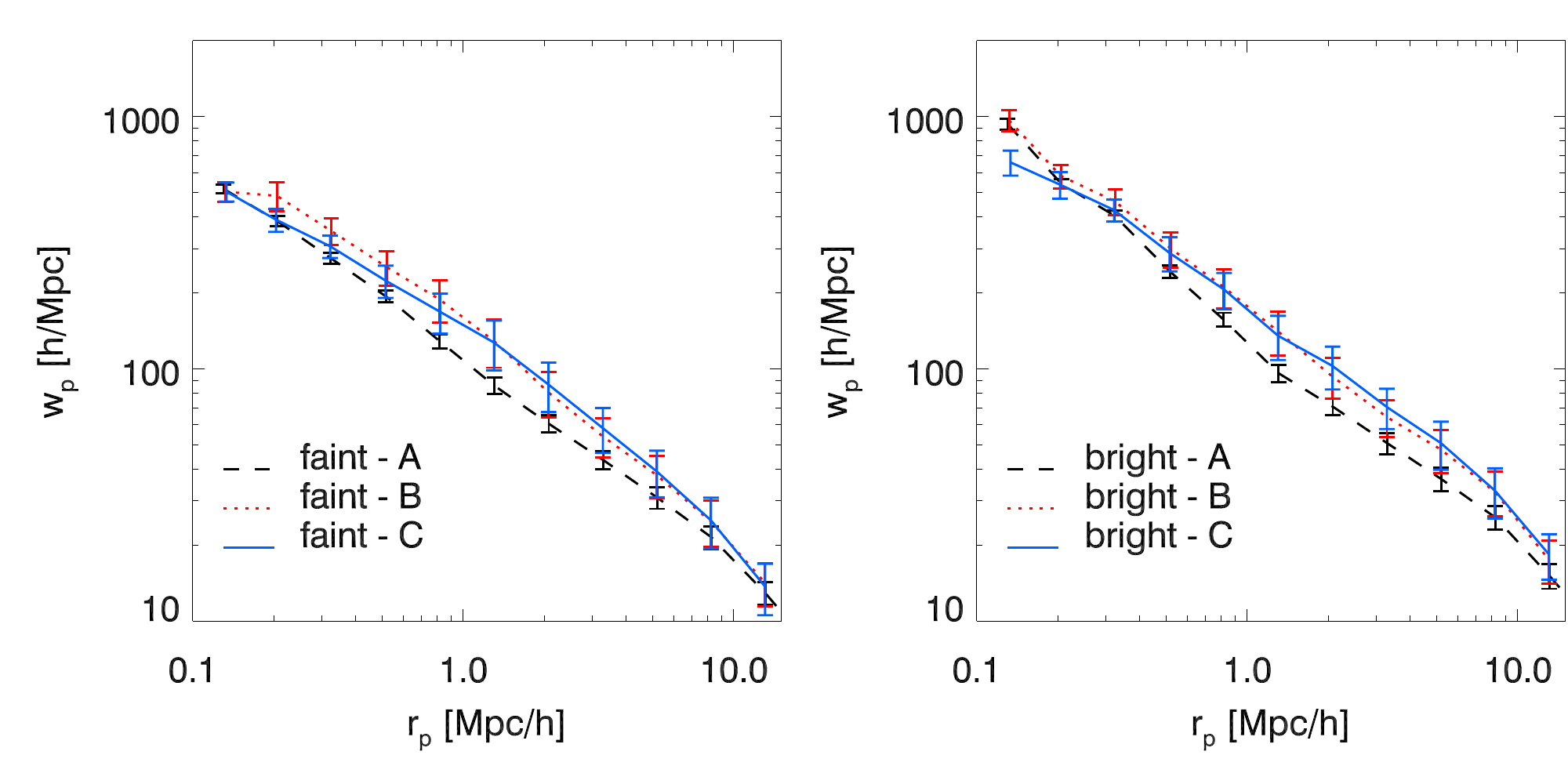}
\caption[Sample variance effects on red galaxies in the Sloan Great Wall]{
cross-correlation functions  of red galaxies in luminosity bin $[-20.5,-21.0]$ for different survey areas. The dashed line are the cross-correlation functions with all $(g-r)>0.85$ galaxies in SDSS in this magnitude bin, the dotted line restricts the SDSS red galaxies to the combined footprint, and the solid line shows the cross-correlation function for $(NUV-r)$ identified red galaxies.}
\label{fig:wpgw}
\end{figure}

\begin{table*}
\begin{center}
\caption{best-fit HOD model parameters for SDSS samples}
\label{tab:HODsdss}
\begin{tabular}{c|cccc|c}
\hline\hline
$\mr{lg}\, M_{\mr r}$ & 
$\mr{lg}\,M_{\mr{min}}^{\mr f}$ & 
$\sigma_{\mr{log}M}^{\mr f}$ & 
$\mr{lg}\, M_0^{\mr f}$ & 
$\mr{lg}\, M_1^{\prime,\mr f}$ & 
$\chi^2/\mr{d.o.f}$\\
\hline
$[-19.5,-20.0]$&  $11.55\pm 0.04 $ &   $ 0.24\pm 0.11   $&  $ 10.14\pm 0.15  $  & $ 12.80\pm 0.03  $&1.80 \\  
$[-20.0,-20.5]$&   $ 11.64\pm 0.02 $  &  $0.16\pm 0.08  $ &  $ 10.07\pm 0.14 $ &   $ 12.92\pm 0.03 $&3.04  \\
$[-20.5,-21.0]$ &  $ 11.98\pm 0.09 $  &  $0.42\pm 0.13 $   &  $  9.45\pm 0.28  $ & $  13.12\pm0.04  $& 3.85 \\  
$[-21.0,-21.5]$&   $12.20\pm0.03  $ &  $0.18\pm 0.07 $   &  $11.77\pm 0.26  $  &  $ 13.45\pm 0.03 $ &  3.84\\ 
\hline
$\mr{lg}\, M_{\mr r}$ & 
$\mr{lg}\, M_{\mr{min}}^{\mr b}$ & 
$\sigma_{\mr{log}M}^{\mr b}$ & 
$\mr{lg}\,  M_0^{\mr b}$ & 
$\mr{lg}\,M_1^{\prime,\mr b}$ & 
$\chi^2/\mr{d.o.f}$\\
\hline
$[-19.5,-20.0]$& $  12.01\pm 0.04  $  &$ 0.26\pm 0.10 $  &$11.08\pm 0.33  $ & $13.27\pm 0.03  $ &  1.80 \\
$[-20.0,-20.5]$ &  $  12.26\pm 0.03 $  & $ 0.36\pm 0.14 $  &$ 11.91\pm 0.10 $  &$ 13.46\pm 0.03 $   &  3.04\\
$[-20.5,-21.0]$& $  12.97\pm 0.08 $  & $ 0.80\pm 0.26 $  &$ 10.59\pm 0.42 $  & $13.81\pm 0.03 $   & 3.85 \\  
$[-21.0,-21.5]$&  $ 13.41\pm 0.04 $  & $ 0.69\pm 0.14 $  & $11.95\pm 0.23 $  & $14.41\pm 0.03 $   & 3.84\\
\hline
\hline
\end{tabular}
\end{center}
\end{table*}
Based on these HOD models for the SDSS reference samples, we now turn to the color selected galaxy samples. Figure~\ref{fig:hodfit} shows the measured cross-correlation functions between color samples and the SDSS reference samples, the best-fit model correlation functions, and the best-fit halo occupation distribution. For comparison, we also show the properly weighted sum of the color sample HOD models, and the best-fit HOD for the color independent sample of SDSS galaxies in the sample luminosity bin. While the characteristic mass scales of these HODs are similar, such comparisons are limited by the large degeneracies between fit parameters\footnote{Ideally, one would fit all three color samples simultaneously and use the sum of the three color sample HODs to fit the correlation functions of the color independent luminosity bin sample. However, the survey area of our current sample is not sufficient to estimate the large covariance matrices required for such an analysis.}. Overall, these models provide acceptable fits to the measured correlation functions, with an exception for the green and red galaxy samples in luminosity bin $[-20.5,-21.0]$. These correlation functions have an unusual flat shape, do not show the characteristic transition from one-halo to two-halo term regime, and the typical host halo masses inferred from the two-halo regime are significantly larger than those inferred from the one-halo term only. As discussed in section \ref{sec:blin}, the redshift of this luminosity bin is centered on the Sloan Great Wall, which is contained almost completely in the angular mask of the SDSS-GALEX cross-match. Hence the clustering measurements in this luminosity bin may be affected by the overdense environment, and as the great Wall occupies a disproportionally large fraction of the combined footprint, the jack knife error bars may underestimate the sample variance. For comparison we show the cross-correlation functions of $(g-r)$ color identified red galaxies in this luminosity bin computed over the full SDSS area and the combined survey footprint in Fig.~\ref{fig:wpgw}. The clustering of $(NUV-r)$ and $(g-r)$ selected red galaxies in the joint survey geometry is nearly indistinguishable, while the cross-correlation function of red galaxies in this luminosity bin over the full SDSS area has the expected shape. It can be fit with a color bin HOD model with reduced $\chi^2 = 3.2$, suggesting that the poor fit in Fig.~\ref{fig:hodfit} is indeed caused by the Great Wall structure and not a systematic effect in the construction of the $(NUV-r)$ selected galaxy sample. 

\begin{figure*}
\includegraphics[width = \textwidth]{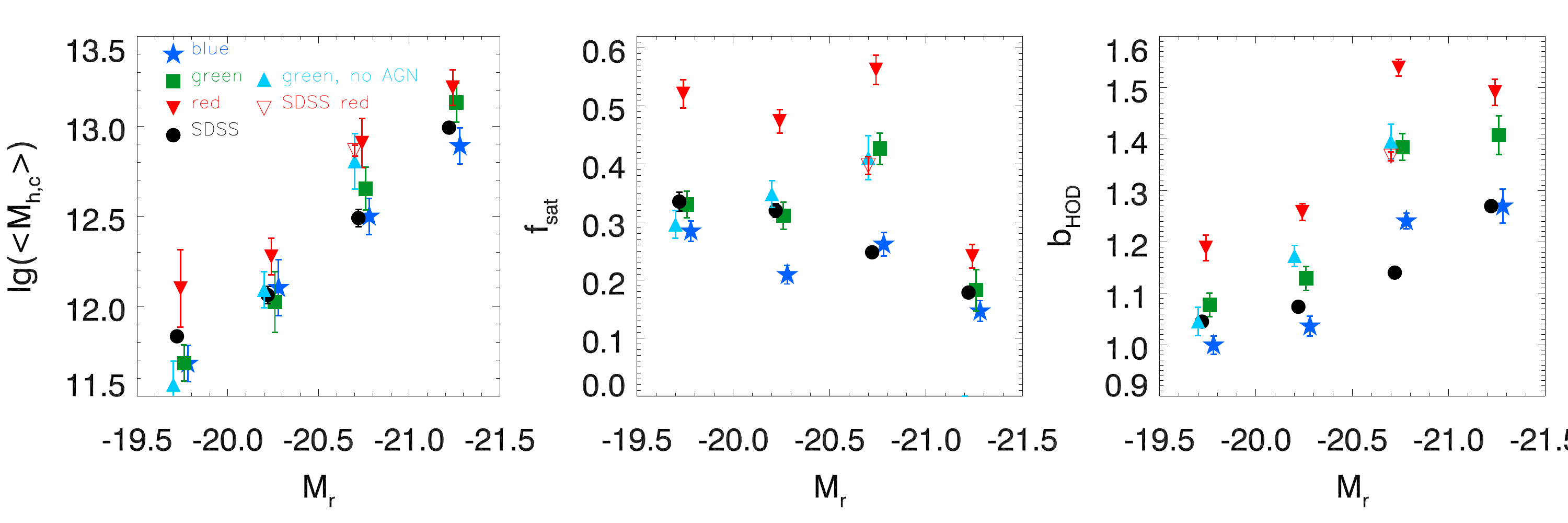}
\caption[Derived HOD parameters for $(NUV-r)$ selected samples]{Derived HOD parameters for luminosity and color bin samples. Left: Mean halo mass for a halo have a central galaxy from a particular sample. Middle: Satellite fraction as a function of galaxy luminosity and color. Right: Galaxy bias derived from the HOD model fit.}
\label{fig:hod1}
\end{figure*}

\begin{table*}
\begin{center}
\caption{best-fit HOD model derived parameters, and their correlation coefficients, $\rho$, for color selected galaxy samples}
\label{tab:HODnuv}
\begin{tabular}{cc|ccc|c|rrr}
\hline\hline
$\mr{lg}\, M_{\mr r}$ & color &
$\mr{lg}\ensav{M_{\mr{h,c}}}$ & $f_{\mr{sat}}$ & $b_{\mr{HOD}}$ &
$\chi^2/\mr{d.o.f}$ &$\rho(\mr{lg}\ensav{M_{\mr{h,c}}},f_{\mr{sat}})$& $\rho(\mr{lg}\ensav{M_{\mr{h,c}}},b_{\mr{HOD}})$&$\rho(f_{\mr{sat}},b_{\mr{HOD}})$ \\
\hline
$[-19.5,-20.0]$&  blue &$11.68\pm 0.10 $ &   $ 0.28\pm 0.02   $&  $ 0.99\pm 0.02  $&1.81 &-0.13&0.45&0.82\\  
$[-19.5,-20.0]$&  green &$11.68\pm 0.12 $ &   $ 0.33\pm 0.03   $&  $ 1.07\pm 0.02  $&3.72&0.05&0.35&0.91 \\  
$[-19.5,-20.0]$&  red &$12.09\pm 0.21 $ &   $ 0.51\pm 0.03   $&  $ 1.18\pm 0.03  $&3.51 &-0.36&0.58&0.54\\  
$[-19.5,-20.0]$&  all &$11.83\pm 0.03 $ &   $ 0.33\pm 0.01   $&  $ 1.04\pm 0.01  $&2.04 &0.21&-0.01&0.09\\  
\hline
$[-20.0,-20.5]$&  blue &$12.10\pm 0.15 $ &   $ 0.21\pm 0.02   $&  $ 1.04\pm 0.02  $&1.20 &-0.60&0.38&0.73\\  
$[-20.0,-20.5]$&  green &$12.02\pm 0.17 $ &   $ 0.31\pm 0.02   $&  $ 1.12\pm 0.02  $&1.02 &-0.49&0.51&0.46\\  
$[-20.0,-20.5]$&  red &$12.28\pm 0.10 $ &   $ 0.47\pm 0.02   $&  $ 1.25\pm 0.01  $&1.04 &-0.39&0.11&0.86\\  
$[-20.0,-20.5]$&  all &$12.06\pm 0.04 $ &   $ 0.31\pm 0.01   $&  $ 1.07\pm 0.01  $&1.30 &-0.73&0.41&0.26\\  
\hline
$[-20.5,-21.0]$&  blue &$12.49\pm 0.11 $ &   $ 0.26\pm 0.02   $&  $ 1.24\pm 0.02  $&2.93 &-0.59&0.47&0.41\\  
$[-20.5,-21.0]$&  green &$12.65\pm 0.12 $ &   $ 0.42\pm 0.03   $&  $ 1.38\pm 0.03  $&5.90 &-0.34&0.46&0.67\\  
$[-20.5,-21.0]$&  red &$12.91\pm 0.14 $ &   $ 0.56\pm 0.03   $&  $ 1.53\pm 0.02  $&12.06 &-0.69&0.35&0.42\\  
$[-20.5,-21.0]$&  all &$12.49\pm 0.05 $ &   $ 0.25\pm 0.01   $&  $ 1.14\pm 0.01  $&2.32 &-0.52&0.77&-0.01\\  
\hline
$[-21.0,-21.5]$&  blue &$12.89\pm 0.09 $ &   $ 0.14\pm 0.01   $&  $ 1.26\pm 0.02  $&2.88 &-0.69&0.81&-0.24\\  
$[-21.0,-21.5]$&  green &$13.13\pm 0.11 $ &   $ 0.18\pm 0.03   $&  $ 1.41\pm 0.04  $&1.54 &-0.71&0.72&-0.08\\  
$[-21.0,-21.5]$&  red &$13.21\pm 0.10 $ &   $ 0.24\pm 0.02   $&  $ 1.49\pm 0.03  $&2.02 &-0.56&0.67&0.13\\  
$[-21.0,-21.5]$&  all &$12.99\pm 0.02 $ &   $ 0.18\pm 0.01   $&  $ 1.27\pm 0.01  $&2.54&-0.65&0.79&-0.12 \\  
\hline
\hline
\end{tabular}
\end{center}
\end{table*}

Figure~\ref{fig:hod1} and Tab.~\ref{tab:HODnuv} show marginalized constraints on the mean mass of halos hosting a central galaxy of given color and luminosity (which is different from $M_{\mr c}^X$, as it also depends on the scatter $\sigma_X$), satellite fraction, and HOD derived galaxy bias for color and luminosity bin samples based on the parameterization described in section \ref{sec:HODnuv}. We show these derived quantities instead of the HOD parameters as they are less affected by degeneracies between the fit parameters, which cause large marginalized errors in the individual fit parameters.

Based on this simple parameterization, we find red central galaxies to occupy more massive halos than the average central galaxy from the same luminosity bin. Within the statistical uncertainty due to the small size of our color selected galaxy samples, there is no significant difference between the halo masses of blue and green central galaxies. At fixed luminosity, the satellite fraction and HOD derived galaxy bias increases with $(NUV-r)$ color. The former is consistent with the results of \citet{Zehavi10} who found the satellite fraction to vary smoothly with $(g-r)$ color at fixed luminosity. This analysis used a one-parameter family of models based on the HOD of the color independent luminosity threshold sample with only the normalization of the satellite galaxy occupation function as a free parameters. Note that given the similarities in central galaxy halo masses, differences in the HOD derived bias parameters mainly reflect the changes in the mean halo mass for satellite galaxies. This implies that the host halo masses of green satellite galaxies are intermediate between those of blue and red satellite galaxies. 

Overall, we find the slope of the satellite occupation distribution,$\alpha_X$, and radial concentration parameter, $c_X$ to increase with $(NUV-r)$ color. However, the degeneracies between HOD parameters are large and do not allow us to put reliable constraints on their luminosity dependence.

For luminosity bin $[-20.5,-21.0]$ we also show results derived from $(g-r)$ selected red galaxies in the full SDSS area to indicate the impact of the Sloan Great Wall. In the Great Wall the satellite fraction and halo mass of red galaxies is increased compared to a more representative survey volume, as expected from the color-density relation. As the $(NUV-r)$ color selected samples in this luminosity bin are subject to increased sample variance, the results for blue and green galaxies in this luminosity bin should similarly be interpreted with caution.

As noted by \citet{Martin07} and \citet{Salim07}, a large fraction of active galactic nuclei (AGN) have green $(NUV-r)$ colors. These galaxies may be transitional galaxies with star formation being quenched by AGN feedback \citep[e.g., after undergoing a major merger,][]{Springel05}, or red sequence interlopers which appear green due to the $NUV$ AGN continuum emission.
We test whether the intermediate clustering of green valley galaxies is caused by AGN, which may be a different population than the non-AGN transitional galaxies. We identify green AGN through emission line diagrams \citep{BPT} using the \citet{Kewley01} extreme starburst classification line. We use the emission line measurements from the MPA-JHU catalog and require a signal-to-noise $S/N\geq3$ in the emission lines. Our goal is to remove any potential AGN contamination from the green valley galaxy sample, and we remove all galaxies which are classified as AGN in at least one of the three diagrams as this allows us to categorize galaxies which do not meet the $S/N$ threshold for all emission line. Repeating our clustering and HOD analysis for non-AGN green galaxies we find the HOD of green non-AGN galaxies to be indistinguishable of that of green galaxies including AGN, in agreement with trends earlier observed by \citep{L06,H09}. We do not show results derived from HOD fits for the non-AGN green valley galaxies in luminosity bin $[-21.0,-21.5]$ as this sample is too small to obtain stable covariances.
\section{Summary and Discussion}
We introduced a new analysis and HOD modeling technique for galaxy cross-correlation functions using multiple tracer populations. This approach is particularly useful for interpreting the clustering of  sparse and/or luminosity bin selected galaxy samples of interest. It is advantageous for the analysis of sparse galaxy samples as considering the cross-correlation function with more abundant galaxy populations significantly reduces the statistical uncertainty. 

While the galaxy number density provides strong constraints on the HOD of luminosity threshold samples, the HOD of luminosity bin samples is independent of the galaxy abundance; in this case considering the cross-correlation with multiple tracer populations is particularly useful as it provides an additional mass scale for the calibration of the luminosity bin HOD. An additional advantage of this method is that modeling the CCF between a color selected sample and a color independent sample does not require assumptions on the correlation between central and satellite colors.

This allows us to constrain the central galaxy HOD of color and luminosity bin selected samples for the first time.
We apply this multiple tracer technique to analyze the clustering of $(NUV-r)$ color selected blue, red, and green valley galaxy samples. Our key result is that halo mass of central galaxies, satellite fraction, and halo mass of satellite galaxies increase with $(NUV-r)$ color at fixed luminosity. 

While our results indicate that the clustering properties of green valley galaxies are consistent with them being an intermediate population between blue and red galaxies, the $(NUV-r)$ selected green valley galaxy samples in this analysis consist of only about one thousand galaxies and are too small to provide insight on the transition mechanism(s) at work. In particular, the HOD parameters which describe the abundance and distribution of color selected satellite galaxies, i.e. the slope of the satellite occupation function, $\alpha_X$, and the color dependence of the radial satellite distribution, $c_X$, are poorly constrained by the data. With data from future galaxy redshift surveys, these parameters will provide information on the efficiency of star formation quenching as a function of halo mass and location within a halo. Furthermore, larger data sets will enable a detailed measurement of the redshift space correlation function and thus enable constraints on the infall stage and satellite orbits of transitional galaxies.

The reduced $\chi^2$ values of the best-fit HODs in our analysis of color selected galaxy samples are relatively large, and our model is particularly insufficient to reproduce the clustering of galaxies in or near the Sloan Great Wall. Overall, it is not surprising that a five parameter HOD model does not fully describe the color dependent clustering of galaxies. While the HOD formalism works well to describe the overall relation between (color independent) galaxies and their halos, it is questionable if the strong assumptions implicit in the HOD formalism, such as the one-to-one relation between halo mass an bias, hold for each sub-population.
Additionally, numerical and observational results indicate that the influence of massive halos may extend beyond $R_{200}$, e.g., through highly eccentric satellite orbits \citep{Benson05, Wetzel11} and infall related shocks extending beyond the virial radius \citep[e.g.,][]{Balogh00}, which is not easily incorporated in halo models. 

Finally we note that \citet{Behroozi10, Leauthaud11} recently proposed an improved HOD parameterization based on a detailed model for the relation between stellar mass and halo mass. Their results (figure 3 in \citet{Leauthaud11}) indicate that halo masses derived from the HOD parameterization for luminosity threshold samples adopted in our analysis (equation (\ref{eq:NM})) may be biased by up to 40\%, with the main source of this discrepancy being the assumptions of a power-law form and constant scatter for the luminosity-halo mass relation. For luminosity bin samples, however, these assumptions are better justified, and we expect only small discrepancies between different HOD parameterizations.
\section*{Acknowledgements}
We thank David Weinberg and Zheng Zheng for helpful discussions about HOD modeling of binned samples.

During the preparation of this work, E.K. and C.M.H. were supported by the US National Science Foundation (AST-0807337) and the David \& Lucile Packard Foundation. C.H. was additionally supported by the US Department of Energy (DE-SC0006624).
Funding for the SDSS and SDSS-II has been provided by the Alfred P. Sloan Foundation, the Participating Institutions, the National Science Foundation, the U.S. Department of Energy, the National Aeronautics and Space Administration, the Japanese Monbukagakusho, the Max Planck Society, and the Higher Education Funding Council for England. The SDSS Web Site is http://www.sdss.org/.

The SDSS is managed by the Astrophysical Research Consortium for the Participating Institutions. The Participating Institutions are the American Museum of Natural History, Astrophysical Institute Potsdam, University of Basel, University of Cambridge, Case Western Reserve University, University of Chicago, Drexel University, Fermilab, the Institute for Advanced Study, the Japan Participation Group, Johns Hopkins University, the Joint Institute for Nuclear Astrophysics, the Kavli Institute for Particle Astrophysics and Cosmology, the Korean Scientist Group, the Chinese Academy of Sciences (LAMOST), Los Alamos National Laboratory, the Max-Planck-Institute for Astronomy (MPIA), the Max-Planck-Institute for Astrophysics (MPA), New Mexico State University, Ohio State University, University of Pittsburgh, University of Portsmouth, Princeton University, the United States Naval Observatory, and the University of Washington.

\label{lastpage}
\end{document}